\documentclass[useAMS,usenatbib,onecolumn]{mn2e}

\usepackage{psfig}

\newcommand{\rd}{{\rm d}}

\title[SPH with flux-limited diffusion]{Smoothed particle hydrodynamics with radiative transfer in the flux-limited diffusion approximation}

\author[S. C. Whitehouse \& M. R. Bate]
  {Stuart C. Whitehouse and
  Matthew R. Bate\thanks{E-mail: scw@astro.ex.ac.uk, mbate@astro.ex.ac.uk}.\\
  School of Physics, University of Exeter, Stocker Road, Exeter EX4 4QL
}

\date{Accepted for publication in MNRAS}

\begin{document}
\maketitle

\begin{abstract} 
We describe the implementation and testing of a smoothed particle 
hydrodynamics (SPH) code that solves the equations of radiation hydrodynamics
in the flux-limited diffusion (FLD) approximation.  The SPH equations of
radiation hydrodynamics for an explicit integration scheme are derived 
and tested.  We also discuss the implementation of an implicit
numerical scheme for solving the radiation equations that allows
the system to be evolved using timesteps much longer than the explicit 
radiation timestep.  The code is tested 
on a variety of one-dimensional radiation hydrodynamics problems including
radiation propagating in an optically thin medium, optically thick and
thin shocks, subcritical and supercritical radiating shocks, and a radiation
dominated shock.  Many of the tests were also performed by 
\citet{TS2001} to test their implementation of a FLD module for 
the ZEUS-2D code.  The SPH code performs at least as well as the ZEUS-2D code 
in these tests.
\end{abstract}


\section{Introduction}

The smoothed particle hydrodynamics (SPH) technique was first introduced
by \citet{L1977} and \citet{GM1977} as a Lagrangian method for 
solving astrophysical fluid dynamics problems.  Over the subsequent 
years, it has been used to investigate a wide range of astronomical
problems including galaxy formation, star formation, stellar collisions,
supernova explosions, and meteor impacts \citep[see][for a review]{M1992}.  
It has also made its way into other fields of science and engineering, 
having been used to model tsunamis, volcanic eruptions, and ballistics.

Many improvements and extensions to the method have been made over the years.
Particle smoothing lengths that varied in space as well as time were 
introduced \citep{E1988,HK1989,BCPB1990} to allow
the spatial resolution of the method to vary with density.  This feature
allows SPH to compete against the more modern adaptive mesh refinement (AMR)
codes.  Gravity, while naturally requiring a computational effort scaling
as the square of the particle number $N$, was reduced to order $N \log{N}$ 
scaling using hierarchical tree structures \citep{HK1989,BCPB1990}.
  Recently, a reformulation of the SPH equations has
allowed both energy and entropy to be conserved simultaneously 
\citep{II2001,SH2002,M2002}.
Finally, many authors have suggested various reformulations
of SPH's artificial viscosity 
\citep{W1981,GM1983,E1988,Balsara,FMHRR1994,WBFTW1996,MM1997}  
or its elimination
in favour of Godunov schemes \citep{I1994,I1999,II2001,Cha,CW2003}.

However, with a few exceptions, SPH is primarily used to model
pure hydrodynamics.  Incorporating magnetic fields into the SPH formalism 
has proven difficult.  Problems encountered include numerical tensional
instabilities, non-conservation of momentum, and/or not enforcing the
divergence of the magnetic field to be zero 
\citep[see the introduction of][]{PM2004a}.  
Recently, a promising new method was presented 
by \citet{PM2004a,PM2004b}  which may at last overcome these problems.

Similarly, SPH has been used for radiation hydrodynamics 
\citep{L1977,B1985,B1986}, but not without encountering problems.  
Boundary conditions that allow appropriate heat losses can be difficult
to implement.  More importantly, the radiation equations involve the second
spatial derivative of the radiation energy density.  In the standard SPH 
formalism, the spatial derivative of a quantity is calculated using a 
sum over neighbouring particles, introducing noise.  Calculating a 
second spatial derivative from such a first derivative produces a result
that is very sensitive to particle disorder.
In addition, for many interesting astrophysical problems, the timestep
required to evolve the radiation equations explicitly is many orders of
magnitude shorter than the dynamical timescale of the problem.  Another
recent approach to solving radiation transport with SPH is to combine SPH 
with Monte-Carlo radiative transfer \citep{OW2003}.  This has
the advantage that the radiative transfer is not carried out using the
SPH formalism; the radiative transfer is calculated between hydrodynamical
timesteps and the resulting temperatures are applied to the SPH particles.  
Since the method involves following photon packets, it is simple to 
include radiating
point sources (e.g. stars) and scattering, and it allows 
frequency-dependent radiative transfer.  Its main disadvantages are 
that it assumes the matter and radiation temperatures are the same and 
it is exceedingly computationally expensive, especially in the optically thick
regime. 

Recently, \citet{CM1999} published a method for modelling
heat conduction with SPH.  The heat conduction equation is similar in
form to the equation for radiative diffusion.  Cleary \& Monaghan's method
avoids the sensitivity to particle disorder by reformulating the heat
conduction equation to use a first-order rather than second-order spatial 
derivative.  This solves one of the hinderances for implementing radiative
transfer within the SPH formalism.  \citet*{JSD2004}
recently used this method to study heat conduction in galaxy clusters.  

In this paper, we describe the implementation of an SPH code that solves
the equations of radiation hydrodynamics in the flux-limited diffusion
approximation.  The flux-limited diffusion approximation is briefly reviewed
in Section 2.  Section 3 describes the SPH implementation in detail.
Section 4 presents the results of various test calculations using the
code.  Our conclusions and some directions for future development 
are summarised in Section 5.

\section{Equations of radiation hydrodynamics}
\label{sec:theory}
In a frame comoving with a radiating fluid, assuming local thermodynamic 
equilibrium (LTE), the coupled equations of radiation hydrodynamics (RHD)
to order unity in $v/c$ are
\begin{equation}
\label{rhd1}
\frac{D\rho}{Dt} + \rho\mbox{\boldmath $\nabla\cdot v$} = 0~,
\end{equation}
\begin{equation}
\label{rhd2}
\rho \frac{D\mbox{\boldmath $v$}}{Dt} = -\nabla p + \frac{\mbox{${\chi_{}}_{\rm \scriptscriptstyle F}\rho$}}{c} \mbox{\boldmath $F$}~,
\end{equation}
\begin{equation}
\label{rhd3}
\rho \frac{D}{Dt}\left( \frac{E}{\rho}\right) = -\mbox{\boldmath $\nabla\cdot F$} - \mbox{\boldmath $\nabla v${\bf :P}} + 4\pi \kappa_{\rm \scriptscriptstyle P} \rho B - c \kappa_{\rm \scriptscriptstyle E} \rho E~,
\end{equation}
\begin{equation}
\label{rhd4}
\rho \frac{D}{Dt}\left( \frac{e}{\rho}\right) = -p \mbox{\boldmath $\nabla\cdot v$} - 4\pi \kappa_{\rm \scriptscriptstyle P} \rho B + c \kappa_{\rm \scriptscriptstyle E} \rho E~,
\end{equation}
\begin{equation}
\label{rhd5}
\frac{\rho}{c^2} \frac{D}{Dt}\left( \frac{\mbox{\boldmath $F$}}{\rho}\right) = -\mbox{\boldmath $\nabla\cdot${\bf P}} - \frac{\mbox{${\chi_{}}_{\rm \scriptscriptstyle F}\rho $}}{c} \mbox{\boldmath $F$}~
\end{equation}
 \citep{MM1984,TS2001}.  In these equations, 
$D/Dt \equiv \partial/\partial t + \mbox{\boldmath $v \cdot \nabla$}$  is 
the convective derivative.  The symbols $\rho$, $e$, {\boldmath $v$}, 
and $p$, represent the material mass density, energy density, 
velocity, and scalar isotropic pressure, respectively.  The total 
frequency-integrated radiation energy density, momentum density or flux,
and pressure tensor are represented by $E$, {\boldmath $F$}, and {\bf P},
respectively.  These are the zeroth, first and second angular moments of
the radiation specific intensity which, in general, varies with spatial
position, viewing direction, frequency, and time.

A full description of the flux-limited diffusion approximation to the 
above equations is given by \citet{TS2001}.  Here we simply
summarise the main points.  The assumption of LTE allows the rate of
emission of radiation from the matter in equations \ref{rhd3} and \ref{rhd4}
to be written as the Planck function, $B$.  Equations \ref{rhd2} to \ref{rhd5}
have been integrated over frequency, leading to the flux mean total opacity
${\chi_{}}_{\rm \scriptscriptstyle F}$, and the Planck mean and energy mean
absorption opacities, $\kappa_{\rm \scriptscriptstyle P}$ and 
$\kappa_{\rm \scriptscriptstyle E}$.  In this paper, the opacities are
assumed to be independent of frequency so that 
$\kappa_{\rm \scriptscriptstyle P}=\kappa_{\rm \scriptscriptstyle E}$
and the subscripts may be omitted.  The total opacity, $\chi$, is the sum
of components due to absorption, $\kappa$, and scattering, $\sigma$.
Note that the above opacities have dimensions of length squared over mass 
(i.e.\ cm$^2$/g) whereas \citet{TS2001} define their opacities to
have dimensions of inverse length (i.e.\ their opacities are equal to
ours multiplied by the density).

The equations of RHD may be closed by an equation of state specifying the
gas pressure, the addition of consitutive relations for the Planck function 
and opacities, and an assumption about the relationship between the angular
moments of the radiation field.  In this paper, we use an ideal equation
of state for the gas pressure $p = (\gamma -1)u\rho$, where $u=e/\rho$ is the 
specific energy of the gas.  Thus, the temperature of the gas is 
$T_{\rm g}=(\gamma -1)\mu u/R_{\rm g}=u/c_{\rm v}$ where $\mu$ is the 
dimensionless 
mean particle mass, $R_{\rm g}$ is the gas constant, and $c_{\rm v}$ is the
specific heat capacity of the gas.  The Planck function 
$B=(\sigma_{\rm \scriptscriptstyle B}/\pi)T_{\rm g}^4$, 
where $\sigma_{\rm \scriptscriptstyle B}$ is the Stefan-Boltzmann constant.
The radiation energy density also has an associated temperature $T_{\rm r}$ 
from the equation $E=4 \sigma_{\rm \scriptscriptstyle B} T_{\rm r}^4/c$.

When the radiation field is isotropic ${\bf \rm P} = \frac{1}{3} E$.  The
Eddington approximation assumes this relation holds everywhere and implies
that, in a steady state, equation \ref{rhd5} becomes
\begin{equation}
\label{eddington}
\mbox{\boldmath $F$} = -\frac{c}{3\chi\rho} \nabla E.
\end{equation}
This expression gives the correct flux in optically thick regions, where
$\chi\rho$ is large.  However, in optically thin regions where 
$\chi\rho \rightarrow 0$ the flux tends to infinity whereas in reality 
$|\mbox{\boldmath $F$}| \le cE$.  Flux-limited diffusion solves this problem
by limiting the flux in optically thin environment to always obey the
above inequality.  \citet{LP1981} wrote the radiation flux
in the form of Fick's law of diffusion as
\begin{equation}
\label{fld1}
\mbox{\boldmath $F$} = -D \nabla E,
\end{equation}
with a diffusion constant given by
\begin{equation}
\label{fld2}
D = \frac{c\lambda}{\chi\rho}.
\end{equation}
The dimensionless function $\lambda(E)$ is called the flux limiter.  The 
radiation pressure tensor may then be written in terms of the radiation
energy density as
\begin{equation}
\label{fld3}
\mbox{\rm \bf P} = \mbox{ \rm \bf f} E,
\end{equation}
where the components of the Eddington tensor, {\bf f}, are given by
\begin{equation}
\label{fld4}
\mbox{\rm \bf f} = \frac{1}{2}(1-f)\mbox{\bf I} + \frac{1}{2}(3f-1)\mbox{\boldmath $\hat{n}\hat{n}$},
\end{equation}
where $\mbox{\boldmath $\hat{n}$}=\nabla E/|\nabla E|$ is the unit vector
in the direction of the radiation energy density gradient and the dimensionless
scalar function $f(E)$ is called the Eddington factor.  The flux limiter
and the Eddington factor are related by
\begin{equation}
\label{fld5}
f = \lambda + \lambda^2 R^2,
\end{equation}
where $R$ is the dimensionless quantity $R = |\nabla E|/(\chi\rho E)$.

Equations \ref{fld1} to \ref{fld5} close the equations of RHD, eliminating
the need to solve equation \ref{rhd5}.  However, we must still choose
an expression for the flux limiter, $\lambda$.  In this paper, we choose
Levermore \& Pomraning's flux limiter 
\begin{equation}
\lambda(R) = \frac{2+R}{6 + 3R + R^2}.
\end{equation}
to allow direct comparison of our results with those of 
Turner \& Stone (2001).  In the optically thin limit
\begin{equation}
\lim_{R\rightarrow \infty} \lambda(R) = \frac{1}{R}
\end{equation}
to first order in $R^{-1}$ so the magnitude of the flux approaches
$|\mbox{\boldmath $F$}| = c |\nabla E|/(\chi\rho R) = c E$, as required.
In the optically thick or diffusion limit
\begin{equation}
\lim_{ R\rightarrow 0}~ \lambda(R) = \frac{1}{3}
\end{equation}
to first order in $R$, so the flux takes the value given by 
equation \ref{eddington}.
Many other forms of flux limiter have been developed to give more
realistic performances on different problems
(see Turner \& Stone 2001 and references therein).

\section{Numerical method}

The radiation terms in equations \ref{rhd2} to \ref{rhd4} are added
to a one-dimensional SPH code that is otherwise standard (e.g. 
\citealp{BCPB1990,M1992}).  The usual hydrodynamical SPH equations will not be
rederived here.  We use the standard cubic spline kernel, $W$, for 
the SPH summations over particles.  The smoothing lengths of particles vary in 
time and space, subject to the constraint that the number 
of neighbours for each particle remains approximately 
constant at $N_{\rm neigh}=8$.
The smoothing length for particle $i$ is given by 
\begin{equation}
h_{ i} = \frac{1}{2} {\rm MAX} \left( | x_{ i} - 
x_{i+5}| + |x_{i} - x_{i+4}|, |x_{ i} - x_{i-5}| + |x_{i} - x_{i-4} |\right),
\end{equation}
 where $x_{i}$
is the position of particle $i$, and $x_{i+4}$ and $x_{i+5}$
are the positions of the fourth and fifth closest particles in the postitive
$x$ direction, and $x_{i-4}$ and $x_{i-5}$ are the positions of the fourth 
and fifth closest particles in the negative $x$ direction.

The SPH equations are integrated using one of two possible integrators:
a predictor-corrector integrator, or a 
second-order Runge-Kutta-Fehlberg integrator.  All particles are advanced
with the same timestep.  The integrators give
essentially identical results, so we only present the results of the
test calculations using one of the integrators, the 
second-order Runge-Kutta-Fehlberg integrator.  We use the standard form of 
artificial viscosity \citep{GM1983,M1992} with 
 strength parameters $\alpha_{\rm_v}=1$ and $\beta_{\rm v}=2$ 
unless otherwise
stated.
The code does not include gravity or magnetic fields.

\subsection{Explicit SPH formulation of the radiation equations}

As SPH is a Lagrangian method, it is natural to use specific energies
for the gas and radiation.  The specific energy of the gas is $u=e/\rho$
and we define the specific radiation energy to be $\xi=E/\rho$.  Equation
\ref{rhd1} does not need to be solved directly since the density of each 
particle is calculated using the standard SPH summation over the particle
and its neighbours.  Equations
\ref{rhd2} to \ref{rhd4} can then be written
\begin{equation}
\label{lagrhd2}
\frac{D\mbox{\boldmath $v$}}{Dt} = -\frac{\nabla p}{\rho}  + \frac{\mbox{$\chi$}}{c} \mbox{\boldmath $F$}~,
\end{equation}
\begin{equation}
\label{lagrhd3}
\frac{D\xi}{Dt} = -\frac{\mbox{\boldmath $\nabla\cdot F$}}{\rho} - \frac{\mbox{\boldmath $\nabla v${\bf :P}}}{\rho} - a c \kappa \left( \frac{\rho \xi}{a} - \left(\frac{u}{c_{\rm v}}\right)^4 \right)~,
\end{equation}
\begin{equation}
\label{lagrhd4}
\frac{Du}{Dt} = -\frac{p \mbox{\boldmath $\nabla\cdot v$}}{\rho} + a c \kappa \left( \frac{\rho \xi}{a} - \left(\frac{u}{c_{\rm v}}\right)^4 \right)~,
\end{equation}
where $a=4 \sigma_{\rm \scriptscriptstyle B}/c$.
The first terms on the right-hand sides of each of equations \ref{lagrhd2} 
and \ref{lagrhd4} are the hydrodynamic terms and are solved in the
usual SPH manner (see below).

The first term on the right-hand side of equation \ref{lagrhd3} is the 
radiation flux term which, as discussed in Section \ref{sec:theory}, is given
by
\begin{equation}
\label{eqn:flux1}
- \frac{\nabla \cdot \mbox{\boldmath $F$}}{\rho} = \frac{1}{\rho} \nabla \cdot \left( { c \lambda \over \kappa \rho } \nabla E \right).
\end{equation}
This has a similar form to the right-hand side of the equation for heat conduction 
\begin{equation}
\label{eqn:conduction}
{\rd u \over \rd t} = \frac{1}{\rho} \nabla \cdot \left( k \nabla T_{\rm m} \right),
\end{equation}
where $k$ is the thermal conductivity.
These equations both involve second-order spatial derivatives.  As mentioned
in the introduction, direct computation of a second-order derivative 
using standard SPH techniques makes the result extremely
sensitive to particle disorder, and may result in unstable integration.
\citet{CM1999} overcome this difficulty by reformulating the second-order
derivative as a first-order derivative using a Taylor series expansion.
A full description of this reformulation is given by \citet{JSD2004}. 
The heat conduction equation in SPH formalism then becomes
\begin{equation}
\label{eq:CM21}
{\rd u_i \over \rd t} = \sum_{j=1}^N { {m_j \over \rho_i \rho_j} \left(
\frac{4 k_i k_j}{k_i + k_j} \right) \left( T_i - T_j \right) {\nabla W_{ij} \over r_{ij}} },
\end{equation}
where $W_{ij} = W(r_{ij},h_{ij})$ with $\mbox{\boldmath $r$}_{ij}=\mbox{\boldmath $r$}_{i}-\mbox{\boldmath $r$}_{j}$ and  $h_{ij}=(h_{i}+h_{j})/2$, and the subscripts denote particles $i$ and $j$.
Hence, the SPH formulation of equation \ref{eqn:flux1} is
\begin{equation}
\label{eqn:SPHflux}
\left({D\xi_{i} \over Dt}\right)_{\rm flux} = \sum_{j=1}^N { m_j \over \rho_i 
\rho_j} c \left[ {4 {\lambda_i \over \kappa_{i} \rho_i} {\lambda_j \over \kappa_{j} \rho_j} \over \left( {\lambda_i \over \kappa_{i} \rho_i} +{\lambda_j \over \kappa_{j} \rho_j} \right) } \right] \left( \rho_{i} \xi_{i} - \rho_{j} \xi_{j}
\right)  {\nabla W_{ij} \over r_{ij}}.
\end{equation}
Because energy is transferred between pairs of particles, this formulation 
ensures that energy is conserved.

The last terms in equations \ref{lagrhd3} and \ref{lagrhd4} control the
interaction between the radiation and the gas.  Note that these terms
are identical except in sign, indicating that energy is conserved. 
They are of the form $T_{\rm r}^4 - T_{\rm g}^4$,
as $\xi = {a T_{\rm r}^4}/ \rho$ 
and $u=c_{\rm v} T_{\rm g}$, similar to that of \citet{BB1975} except for 
our use of specific energies rather than energy densities.
 
In one dimension, the radiation pressure term in equation \ref{lagrhd4} becomes
\begin{equation}
\left({D\xi_{i} \over Dt}\right)_{\rm radpres} = - \frac{1}{\rho} \mbox{\boldmath $\nabla v$} : {\bf P} = 
- \left( \mbox{\boldmath $\nabla \cdot v$} \right)_{i} f_i  \xi_{ i},
\end{equation}
where the Eddington factor is 
\begin{equation}
f_{i}= \lambda_i + \lambda_i^2 \left( { | \nabla 
\left( \rho_i \xi_i \right) | \over \kappa_{i} \rho_i^2 \xi_i } \right)^2. 
\end{equation}
 
The final SPH formulations of equations \ref{lagrhd3} and \ref{lagrhd4} are
\begin{equation}
\label{eqn:SPHRTE}
{D \xi_i \over D t} = - \left( \mbox{\boldmath $\nabla \cdot v$} \right)_i f_i \xi_i + \sum_{j=1}^N { m_j \over \rho_i \rho_j} c 
\left[ {4 {\lambda_i \over \kappa_{i} \rho_i} {\lambda_j \over \kappa_{j} \rho_j} \over \left( {\lambda_i \over \kappa_{i} \rho_i} +{\lambda_j \over \kappa_{j} \rho_j} \right) } \right]
\left( \rho_i \xi_i - \rho_j \xi_j \right)  {\nabla W_{ij} \over
  r_{ij}} - a c \kappa_{i} \left({\rho_i
  \xi_i \over a} - \left( {u_i \over c_{{\rm v},i} } \right)^4 \right),
\end{equation}
\begin{equation}
\label{eqn:SPHRTU}
{D u_i \over D t} = \frac{1}{2} \sum_{j=1}^N \left( {p_i \over
  \rho_i^2 } + {p_j \over \rho_j^2} + \Pi_{ij} \right) m_j
  \mbox{\boldmath $v$}_{ij} \cdot \nabla W_{ij} + a c \kappa_{i}
  \left( {\rho_i \xi_i \over a} - \left( {u_i \over c_{{\rm v},i} } 
\right)^4 \right),
\end{equation}
where $\mbox{\boldmath $v$}_{ij}=\mbox{\boldmath $v$}_{i}-\mbox{\boldmath $v$}_{j}$, and the first term in equation \ref{eqn:SPHRTU} is the 
standard SPH expression for the hydrodynamical energy term 
$p(\mbox{\boldmath $\nabla \cdot v$})/\rho$ when the thermodynamic variable of
integration is energy.  We note that the radiation terms can easily be 
expressed in terms of entropy if one wishes to add them to the SPH formalism
of \citet{SH2002}.
We use the standard SPH artificial viscosity
 
\[ \Pi_{ij} =  \left\{ \begin{array}{ll}
         \left( - \alpha_{\rm v} c_{\rm s} \mbox{ $\mu_{ij}$} + \beta_{\rm v} \mbox{ $\mu_{ij}$}^2 \right) / \rho_{ij}
 & \mbox{if  \boldmath $v$} \mbox{$_{ij} \cdot$}\mbox{\boldmath{ $r$}} \mbox{$_{ij} \leq  0 $  , and } \\
        0  & \mbox{if  \boldmath $v$} \mbox{$_{ij} \cdot$}\mbox{\boldmath{ $r$}} \mbox{$_{ij} > 0$ }. \\
\end{array} \right. \] 
where $\mbox{ $\mu_{ij}$} = { h \left( \mbox{\boldmath $v$}_i - \mbox{\boldmath $v$}_j \right) \cdot \left( \mbox{\boldmath $r$}_i - \mbox{\boldmath $r$}_j \right)
/ \left( \left| \mbox{\boldmath $r$}_i - \mbox{\boldmath $r$}_j \right|^2 + \eta^2 \right)}$, with $\eta^2 = 0.01 h^2$ to prevent numerical divergences if particles get too close together.

The full SPH expression of the momentum equation \ref{lagrhd2} is
\begin{equation}
{D \mbox{\boldmath $v_{i}$} \over D t}  = - \sum_{j=1}^{N} m_{j} \left( {p_{i} \over \rho_{i}^2}
+ { p_{j} \over \rho_{j}^2} + \Pi_{ij} \right) \nabla W(r_{ij},h_{ij}) - 
\frac{\lambda_{i}}{\rho_{i}} \sum_{j=1}^N m_j \xi_{j} \nabla W(r_{ij},h_{ij})~,
\end{equation}
where the first term is the standard hydrodynamical term and the second term is
due to radiation pressure.

\subsubsection{Explicit timestep criteria}

Explicit integration requires that the timesteps obey the Courant conditions
for both the hydrodynamic and radiation processes.  The usual hydrodynamical
SPH timestep criteria are
\begin{equation}
\label{tshydro1}
\rd t_{{\rm Courant},i} = { \zeta h_{i} \over c_{\rm s} + h_{i} \left| \nabla \cdot \mbox{\boldmath $v$} \right|_{i} + 1.2 
\left( \alpha_{\rm v} c_{\rm s} + \beta_{\rm v} h_{i} \left| \nabla \cdot \mbox{\boldmath $v$} \right|_{i} \right) },
\end{equation}
and 
\begin{equation}
\label{tshydro2}
\rd t_{{\rm force},i} = \zeta \sqrt{{h_{i} \over \left| \mbox{\boldmath $a$}_{i}  \right|}},
\end{equation}
where we use a Courant number of $\zeta=0.3$ and 
\mbox{\boldmath $a$}$_i$ is the 
acceleration of particle $i$. With radiation hydrodynamics, we modify
equation \ref{tshydro1} so that the sound speed $c_{\rm s}$ used for 
computing the
timestep includes the contribution from radiation pressure, and is given by
$c_{\rm s} = \left[ {\rm max} \left( \gamma , \frac{4}{3} \right) \frac{P_{\rm tot}}{\rho} \right]^{\frac{1}{2}}$, where $P_{\rm tot}$ is the sum of both
the gas and the magnitude of the 
radiation pressure, except where the sound speed is used as
part of the viscosity, where it remains based on just the gas pressure.
The radiation terms in the energy and momentum equations require additional
constraints.  The flux diffusion timestep is
\begin{equation}
\label{eqn:dtrad}
\rd t_{{\rm rad},i} = \zeta { h_i^2 \rho_{i} \kappa_{i} \over c
  \lambda_{i}}.
\end{equation}
The timestep associated with the radiation pressure term in equation
\ref{eqn:SPHRTE} is simply
\begin{equation}
\rd t_{{\rm radpres},i} = \zeta {1 \over f_{i} \rho_{i} \xi_{i}
  \left( \nabla \cdot v \right)_{i}}.
\end{equation}
The timestep for the effect of radiation pressure on the momentum equation
is taken care of already by the standard SPH force timestep criterion.
The two timesteps for the interaction between radiation and matter are 
\begin{equation}
\label{tsint1}
\rd t_{{\rm int \xi},i} = \zeta {  \xi_{i} \over  a c \kappa_{
    i} \left|  {\rho_{i} \xi_{i} \over a} -  \left(
    {u_{i} \over c_{{\rm v},i} } \right)^4 \right| },
\end{equation}
and
\begin{equation}
\label{tsint2}
\rd t_{{\rm intU},i} = \zeta { u_{i} \over  a c \kappa_{i} \left| 
  {\rho_{i} \xi_{i} \over a} -  \left( {u_{i} \over c_{{\rm v},i} }
 \right)^4 \right| }.
\end{equation}

In cases where particles start off in thermal equilibrium ($T_{\rm g}=
T_{\rm r}$), we found it was sometimes necessary to limit the timesteps 
further to ensure that too large a timestep was not used initially. In 
these cases, we also limited the timestep using $ \rd 
t_{{\rm int \xi},i} = \zeta {  \xi_{i} / \left(  a c  \kappa_{
    i} {\rho_{i} \xi_{i} \over a}  \right) }$, $
\rd t_{{\rm int \xi},i} = \zeta { \xi_{i} / \left( a c   \kappa_{i}
 \left( {u_{i} \over c_{{\rm v},i} } \right)^4 \right) }$, $
\rd t_{{\rm intU},i} = \zeta { u_{i} / \left(  a c \kappa_{i}
  {\rho_{i} \xi_{i} \over a}  \right) }$ and $\rd t_{{\rm intU},i} 
= \zeta { u_{i} / \left(  a c \kappa_{i} \left( {u_{i} 
\over c_{{\rm v},i} } \right)^4 \right) }$.

\subsection{Implicit integration}

Because the speed of light is typically much larger than the hydrodynamical 
velocity, the timestep obtained from the Courant condition 
for these radiative processes is much smaller than that for 
the pure hydrodynamics.  It is therefore desirable to integrate the
radiative transfer implicitly so that a much larger timestep can be 
taken.

\citet{M1997} described an implicit integration scheme for
handling the interaction between dust and gas particles in an SPH code.
This is an iterative method that involves sweeping over all pairs of 
neighbouring particles a number of times.
In Monaghan's case, the drag term between gas and dust particles was 
integrated implicitly.  Here, the radiation and gas energy equations 
are integrated implicitly.  Monaghan tried both backward Euler and
two-step implicit integration methods \citet[section ~4]{M1997}.  We
tried both of these methods and the standard trapezoidal method.  We
found the backward Euler and trapezoidal methods gave similar results,
but that the two-step method did not converge in the limit of a large
number of sweeps.  Since the trapezoidal method is the most
accurate A-stable second-order linear multistep method \citep{D1963}, 
we chose to use this scheme.

For the integration between a time $t=n$ and $t=n+1$
 this scheme states that for
a variable $A(t)$,
\begin{equation}
A_{i}^{n+1} = A_{i}^{n} + {\rd t \over 2} \left( {\rd A_{i}^{n} \over \rd t} + {\rd A_{i}^{n+1} \over \rd t} \right),
\end{equation}
where $A^n$ and $A^{n+1}$ indicate the value of $A$ at times $t=n$ and
 $t=n+1$, respectively.

We also tried an iterative scheme based on the alternating-direction implicit 
(ADI) scheme used by \citet{BB1975} and \citet{TS2001}, where progressively
larger pseudo-timesteps are used between sweeps. However, testing showed that
it was slower than the implicit method used below.

\subsubsection{Implicit flux diffusion}

Implicit integration of the radiation flux term can be performed by considering
the interaction between two SPH particles $i$ and $j$. According to the
trapezoidal scheme
\begin{equation}
\label{eqn:E1iflux}
\xi_{i}^{n+1} = \xi_{i}^n + \frac{1}{2} \frac{\rd t}{N} \left[ {m_{j} \over \rho_{i} 
\rho_{j}} b \left( 
\rho_{i} \xi_{i}^n - \rho _{j} \xi_{j}^n + \rho_{i} \xi_{i}^{n+1} - \rho_{j} 
\xi_{j}^{n+1} \right) \frac{\nabla W_{ij}}{{r}_{ij}}  \right] 
\end{equation}
where $b=\left[ {4 {\lambda_{i} \over \kappa_{i} \rho_{i}} {\lambda_{j} \over \kappa_{j} \rho_{j}} \over \left( {\lambda_{i} \over \kappa_{i} \rho_{i}} +{\lambda_{j} \over \kappa_{j} \rho_{j}} \right) } \right]$ for brevity, and $N$ is the total number of sweeps. 
Similarly for particle $j$
\begin{equation}
\label{eqn:E1jflux}
\xi_{j}^{n+1} = \xi_{j}^n + \frac{1}{2} \frac{\rd t}{N} \left[ {m_{i} \over \rho_{i} \rho_{j} 
} b \left( 
\rho_{j} \xi_{j}^n - \rho_{i} \xi_{i}^n + \rho_{j} \xi_{j}^{n+1} - \rho_{i} 
\xi_{i}^{n+1} \right) \frac{\nabla W_{ij}}{{r}_{ij}} \right].
\end{equation}
It is possible to solve this set of equations for the quantity $\rho_{i} \xi_{i}^{n+1} - 
\rho_{j} \xi_{j}^{n+1}$. The solution is
\begin{equation}
\rho_{i} \xi^{n+1}_{i} - \rho_{j} \xi^{n+1}_{j}  = \left( \rho_{i} \xi^n_{i} - \rho_{j} \xi^n_{j} \right)
\left( { 1 + \frac{1}{2} \frac{\rd t}{N} b \frac{\nabla W_{ij}}{{r}_{ij}} \left( {m_{i} \over \rho_{i} } + {m_{j} \over \rho_{j}} \right)} \over { 1 - \frac{1}{2} \frac{\rd t}{N} b \frac{\nabla W_{ij}}{{r}_{ij}} \left( {m_{i} \over \rho_{i} } + {m_{j} \over \rho_{j}} \right)} \right).
\end{equation}
This is then substituted into equations \ref{eqn:E1iflux} and 
\ref{eqn:E1jflux} to provide the new values of the specific radiation 
energies of 
particles $i$ and $j$.  Performing a sweep involves calculating this
interaction between all particle pairs. We tried both updating the energies
after every interaction, and saving the changes from each interaction and 
updating the energies with them after all pairs had been considered. There was
no significant difference between the two methods; we used the latter method
since it minimizes truncation error.

\subsubsection{Implicit pressure and viscous force}

The $p \nabla \cdot v$ and viscous contribution to the matter energy equation
should also be done implicitly. Using the same technique as above, the 
interaction of particle $j$ with particle $i$ is
\begin{equation}
\label{eqn:U1ipdv}
u_{i}^{n+1} = u_{i}^n  + \frac{1}{2} \frac{\rd t}{N} \left[ \frac{1}{2} \left( { \left( \gamma - 1 \right) u^n_{i} \over \rho_{i} } + { \left( \gamma - 1 \right) u^n_{j} \over \rho_{j} } + \Pi_{ij} \right) m_{j} v_{ij} \nabla W_{ij} + \frac{1}{2} \left( { \left( \gamma - 1 \right) u^{n+1}_{i} \over \rho_{i} } + { \left( \gamma - 1 \right) u^{n+1}_{j} \over \rho_{j} } + \Pi_{ij} \right) m_{j} v_{ij} \nabla W_{ij} \right],
\end{equation}
and vice versa for the effect of $i$ on $j$. Solving
for ${u^{n+1}_{i}/\rho_{i}} + {u_{j}^{n+1}/\rho_{j}}$, we obtain

\begin{eqnarray}
 \left({u_{i}^{n+1} \over \rho_{i}}  +  {u_{j}^{n+1} \over \rho_{j}} \right) & = & 
\left({u_{i}^{n} \over \rho_{i}} + {u_{j}^{n} \over \rho_{j}} \right) +   \\ \nonumber 
& ~ & \frac{\rd t}{N \rho_{i}}  \left[ \frac{1}{2} \Pi_{ij} v_{ij} m_{j} \nabla
W_{ij} + \frac{\left( \gamma - 1 \right)}{4} \left( \frac{ u_{i}^n }{\rho_{i}} +  
 \frac{ u_{j}^n }{\rho_{j}} \right) m_{j } v_{ij} \nabla W_{ij} + 
\frac{\left( \gamma - 1 \right)}{4} \left( \frac{ u_{i}^{n+1} }{\rho_{i}} +  
 \frac{ u_{j}^{n+1} }{\rho_{j}} \right) m_{j } v_{ij} \nabla W_{ij} \right]  + \\ \nonumber
& ~ &  \frac{\rd t}{N \rho_{j}} \left[ \frac{1}{2} \Pi_{ji} v_{ji} m_{i} \nabla
W_{ji} + \frac{\left( \gamma - 1 \right)}{4} \left( \frac{ u_{j}^n }{\rho_{j}} +  
 \frac{ u_{i}^n }{\rho_{i}} \right) m_{i } v_{ji} \nabla W_{ji} + 
\frac{\left( \gamma - 1 \right)}{4} \left( \frac{ u_{j}^{n+1} }{\rho_{j}} +  
 \frac{ u_{i}^{n+1} }{\rho_{i}} \right) m_{i } v_{ji} \nabla W_{ji} \right] .
\end{eqnarray}
Noting that $v_{ij} \nabla W_{ij} = v_{ji} \nabla W_{ji} $, and that
$\Pi_{ij} = \Pi_{ji}$, this can be solved by defining four quantities:
\begin{eqnarray}
c_1 =& \displaystyle \frac{u_{i}^n}{\rho_{i}} + \frac{\rd t}{2 N \rho_{i}} \Pi_{ij} v_{ij} m_{j} \nabla W_{ij} + \frac{\rd t}{N} {\left( \gamma - 1 \right) \over 4 \rho_{i} } \left( {u^n_{i} \over \rho_{i}} + {u_{j}^n \over \rho_{j}} \right) m_{j} v_{ij} \nabla W_{ij}, \nonumber \\
c_2 =& \displaystyle \frac{u_{j}^n}{\rho_{j}} + \frac{\rd t}{2 N \rho_{j}} \Pi_{ij} v_{ij} m_{i} \nabla W_{ij} + \frac{\rd t}{N} {\left( \gamma - 1 \right) \over 4 \rho_{j} } \left( {u^n_{i} \over \rho_{i}} + {u_{j}^n \over \rho_{j}} \right) m_{i} v_{ij} \nabla W_{ij}, \nonumber \\ 
c_3 =& \displaystyle \frac{\rd t}{N} {\left( \gamma - 1 \right) \over 4 \rho_{i} } m_{j} v_{ij} \nabla W_{ij}, \nonumber \\ 
c_4 =& \displaystyle \frac{\rd t}{N} {\left( \gamma - 1 \right) \over 4 \rho_{j} } m_{i} v_{ij} \nabla W_{ij}, \nonumber 
\end{eqnarray}
and the solution for $ {u^{n+1}_{i}/ \rho_{i}} + {u_{j}^{n+1}/ \rho_{j}}$ is
simply
\begin{equation}
{u^{n+1}_{i}/ \rho_{i}} + {u_{j}^{n+1}/ \rho_{j}} = { c_1 + c_2 \over 1 - c_3 - c_4 }.
\end{equation}
This quantity is then substituted into equation \ref{eqn:U1ipdv} and its 
counterpart for $j$ to provide new values of $u_{i}$ and $u_{j}$. 
Note that in fact the viscosity has a dependence on $u$, as
the SPH viscosity involves the sound speed $c_{\rm s} \propto \sqrt{u}$. 
However, we neglect this weak dependence during a sweep in order to simplify 
the mathematics, and simply update the sound speed in advance of the
next sweep. We have not noticed any significant effect of this small 
approximation.

\subsubsection{Implicit radiation pressure}

The radiation pressure term in equation \ref{eqn:SPHRTE} merely involves the divergence of the velocity, and can easily be implicitly integrated thus:
\begin{equation}
\xi_{i}^{n+1} = \xi_{i}^n - \frac{1}{2} \frac{\rd t}{N} \left( \mbox{ \boldmath $\nabla \cdot v$} \right)_{i} \left( \xi_{i}^{n+1} + \xi_{i}^n \right)  f_{i}, 
\end{equation}
whose solution is
\begin{equation}
\xi_{i}^{n+1} = \xi_{i}^n \left( { \left( 1 - \frac{1}{2} \frac{\rd t}{N} \left( \mbox{ \boldmath $\nabla \cdot v$} \right)_{i}  f_{i}  \right) \over \left( 1 + \frac{1}{2} \frac{\rd t}{N} \left(\mbox{ \boldmath $\nabla \cdot v$} \right)_{i} f_{i} \right) } \right).
\end{equation}

\subsubsection{Implicit radiation-matter interaction terms}

The implicit solution of the interaction terms between $\xi$ and $u$ for a
given particle $i$ involves the solution of a quartic (fourth-order polynomial)
equation, which is non-trivial. 
The equations for $\xi$ and $u$ are 
\begin{equation}
\label{eqn:quarticE}
\xi^{n+1}_{i} = \xi^n_{i} - \frac{1}{2} \frac{\rd t}{N} a c  
\left(   \frac{\rho_{i} \kappa_{i}}{a} \xi^n_{i} - 
\frac{\kappa_{i}}{c_{{\rm v},i}^4} \left( u^n_{i} \right)^4 
+  \frac{\rho_{i} \kappa_{i}}{a} \xi^{n+1}_{i} - 
\frac{\kappa_{i}}{c_{{\rm v},i}^4} \left( u^{n+1}_{i} \right)^4 
\right)
\end{equation}
and
\begin{equation}
\label{eqn:quarticU}
u^{n+1}_{i} = u^n_{i} + \frac{1}{2} \frac{\rd t}{N} a c  
\left(   \frac{\rho_{i} \kappa_{i}}{a} \xi^n_{i} - 
\frac{\kappa_{i}}{c_{{\rm v},i}^4} \left( u^n_{i} \right)^4 
+  \frac{\rho_{i} \kappa_{i}}{a} \xi^{n+1}_{i} - 
\frac{\kappa_{i}}{c_{{\rm v},i}^4} \left( u^{n+1}_{i} \right)^4 
\right).
\end{equation}
Here the term to solve for is 
\begin{equation}
X =\rho_{i}  \xi_{i}^{n+1} - \frac{a}{\kappa_{i}} \frac{\kappa_{i}}{ c_{{\rm v},i}^4} \left( u_{i}^{n+1} \right)^4. 
\end{equation}
Defining the quantities
\begin{eqnarray}
c_1 =& \displaystyle \rho_{i} \xi_{i}^n - \frac{1}{2} \frac{\rd t}{N} a c \rho_{i} \left( \frac{ \rho_{i} \kappa_{i}}{a} \xi_{i}^n - \frac{\kappa_{i}}{c_{{\rm v},i}^4} \left( u_{i}^n \right)^4 \right), \nonumber \\
c_2 =& \displaystyle {\kappa_{i} a \over c_{{\rm v},i}^4  \kappa_{i}}, \nonumber \\ 
c_3 =& \displaystyle u_{i}^n + \frac{1}{2} \frac{\rd t}{N} a c \left( \frac{ \rho_{i} \kappa_{i}}{a} \xi_{i}^n - \frac{\kappa_{i}}{c_{{\rm v},i}^4} \left( u_{i}^n \right)^4 \right), \nonumber \\ 
c_4 =& \displaystyle \frac{1}{2} \frac{\rd t}{N} a c \rho_{i} \frac{\kappa_{i}}{a}, \nonumber \\
c_5 =& \displaystyle \frac{1}{2} \frac{\rd t}{N} a c \frac{\kappa_{i}}{a}, \nonumber \\
\end{eqnarray}
the equation to be solved becomes
\begin{equation}
\label{eqn:quartic}
X^4 + { 4 c_3 \over c_5} X^3 + { 6 c_3^2 \over c_5^2} X^2 + \left( {4 c_2 c_3^3
\over c_5^2} + {c_4 \over c_2 c_5^4 } + {1 \over c_2 c_5^4} \right) X + 
{ \left( {c_2 c_3^4 - c_1 }\right) \over c_2 c_5^4 } = 0,
\end{equation}
which can be solved using any method for solving quartic equations (see
Appendix \ref{sec:quartsolve} for the analytical method).
While it is possible to use a numeric solver, e.g. a Newton-Raphson method, 
we have found that the analytic solution is to be preferred since
tests show it is much quicker.  One minor difficulty when calculating the 
analytic solution is that while two of the roots of a quartic equation 
(usually) contain an imaginary component and can readily be discarded, 
the code must be able to 
distinguish between the two real roots. We choose 
the correct
solution as the one which gives a new gas temperature that lies between 
the old values of $T_{\rm g}$ and $T_{\rm r}$. 
In the case of four real roots, the same test
can be applied. Four complex roots usually indicates that the timestep 
is too large.

If the real root chosen is denoted by $X_1$, the quantity $ {\kappa_{i} \over a } X_1 $ should be substituted back into equations \ref{eqn:quarticE} and 
\ref{eqn:quarticU} as the term $\left( \frac{\rho_{i} \kappa_{i}}{a} \xi^{n+1}_{i} - 
\frac{\kappa_{i}}{c_{{\rm v},i}^4} \left( u^{n+1}_{i} \right)^4 
\right)$ to provide the new radiation and gas energies.   We have found that
solving for $X_1$ as $\rho_{i} \xi_{i}^{n+1} - \frac{a}{\kappa_{i}} \frac{\kappa_{i}}{c_{{\rm v},i}^4} \left( u_{i}^{n+1} \right)^4 $ rather than $\left( \frac{\kappa_{i}}{a} \rho_{i}  \xi^{n+1}_{i} - 
\frac{\kappa_{i}}{c_{{\rm v},i}^4} \left( u^{n+1}_{i} \right)^4 
\right)$
can be more accurate if the numbers involved in the calculation are closer to 
the order of unity, ensuring that the range does not fall outside the Fortran 
{\sc double precision} variable bounds. For different values of opacity it
may even be desirable to solve for $ \frac{a}{\kappa} X$ or even $\left( \frac{a}{\kappa} \right)^2 X$ depending on the problem. The code can be 
made adaptive in this respect by checking that the co-efficient of the 
first-order term in equation \ref{eqn:quartic} is smaller than the fourth
root of the largest allowed floating point number, and adapting the 
quantity to solve for accordingly. 

The analytical solution of this quartic equation, whilst being much faster
than a numerical solver, can fail when the number of sweeps is too small. 
In this case,
the code detects a failure to reach a physical temperature and returns to the
calling subroutine requesting a greater number of sweeps. This is a fairly
common occurance when the code is just starting, as the initial number of 
sweeps is an estimate input by the user.

\subsubsection{The Adaptive Sweeping Scheme}

The way the sweeping is performed is important, and can greatly affect
the accuracy of the implicit method.  Initially 
the quantities $\xi$ and $u$ are copied into arrays upon which 
the sweeping
operations are to be performed. Then sweeping commences in the following order:
First the flux-limiter $\lambda_{i}$ is
calculated for each particle, along with the Eddington factor $f_{i}$. 
The next step is to calculate the flux diffusion term for all $i-j$ 
particle pairs, update $\xi$, then calculate the radiation pressure term and
update $\xi$ again.  
Independent of the
radiation energy calculations, 
the sound speeds $c_{\rm s}$ (used in the SPH viscosity) are calculated, 
followed by the gas pressure/viscosity term and $u$ is updated. 
The sweep ends with the 
interaction
term between $u$ and $\xi$, after which the next sweep commences with the 
calculation of the flux-limiter using the new values of $\xi$ and $u$. 
Once the specified number of sweeps are complete the temporary arrays are
copied back into their normal counterparts. 

Since it is not known {\it a priori} how many sweeps are required to reach
a certain accuracy, we use an Adaptive Sweeping Scheme (ASS) 
to determine the necessary number of sweeps. The code begins by trying an 
initial number of sweeps $2^m$ (where $m$ is an integer, $ m > 0$).
The implicit subroutine is called
twice, once for $2^{m-1}$ sweeps and once for $2^{m}$ sweeps. If the fractional
 error
in both $\xi$ and $u$ for all particles between the two sweeps is less 
than a specified tolerance the values of $\xi$ and $u$ from the $2^{m}$ 
sweep calculation are used, and $m$ is decreased by unity for the next 
timestep to allow the code to adapt if fewer sweeps are required.  
We found that a tolerance of $10^{-3}$ gave a good trade off between 
accuracy and speed and this value is used for all the test calculations
presented in this paper.
If the error is above the tolerance, $m$ is incremented by unity, 
and the method repeats until either an acceptable solution is found, or
 too many
sweeps are attempted. The trapezoidal scheme is such that doubling the 
number of sweeps typically halves the errors. The ASS can be optimized 
to take this into account, e.g. if the maximum error is one quarter of the
acceptable tolerance the number of sweeps can be reduced by a factor of four 
for the next timestep. If at any time the implicit
integration would take $\xi$ or $u$ negative, the implicit routine immeditately
returns to ASS with a message to double the number of sweeps, 
similar to what occurs when the quartic solver fails.

It is also desirable for two-step integrators like the Runge-Kutta-Fehlberg 
or predictor-corrector integrators that require the calculation of 
quantities at both
the half and the full timesteps to use two independent numbers of 
sweeps.  Typically, the full timestep call of the implicit method requires 
twice as many sweeps.

\section{Test calculations}

\subsection{Heating and cooling terms}

\label{sec:heatcool}

As in \citet{TS2001} we tested the interaction between the radiation and the 
gas to check that the temperatures of the gas and the radiation equalise
when $T_{\rm g} \neq T_{\rm r}$ initially.  A gas 
was set up so that there was no initial velocity, with a density
$\rho = 10^{-7}$ g cm$^{-3}$, opacity $\kappa = 0.4$ cm$^2$ g$^{-1}$, and 
$\gamma=
 \frac{5}{3}$ and  $E = 10^{12}$ ergs cm$^{-3}$. 
Two tests were carried out, one where the gas heated until it
reached the radiation temperature, and one where it cooled. The first test had
$e=u \rho = 10^2$ ergs cm$^{-3}$, and the second 
$e=u \rho = 10^{10}$ ergs cm$^{-3}$. 
The boundaries of the calculation used reflective ghost particles, and the
position of the boundary was fixed.

\begin{figure}
\centerline{\psfig{figure=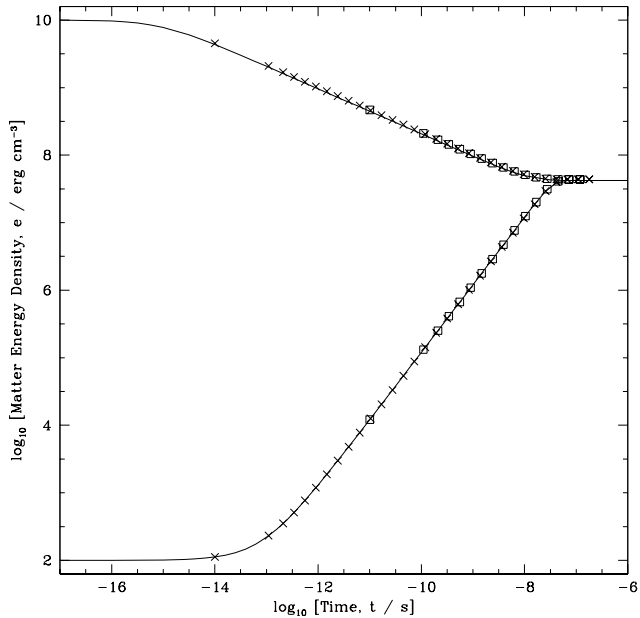,width=9.0truecm}}
\caption{\label{fig:ts51} 
The evolution of the gas energy density $e$ as it equilibriates with a 
radiation energy density $E=10^{12}$ erg cm$^{-3}$.  In the upper case, 
$e=10^{10}$ erg cm$^{-3}$ initially, while in the lower case
$e=10^{2}$ erg cm$^{-3}$. The
solid line is the analytic solution, the crosses are the results of the SPH
code using implicit timesteps of the greater of $10^{-14}$~s or five percent 
of the 
elapsed time, and the squares with a timestep of the larger of $10^{-11}$~s or
five percent of the elapsed time. The symbols are plotted every ten timesteps.}
\end{figure}

This problem can be approximated by the differential equation\begin{equation}
\frac{\rd e}{\rd t} = c \kappa E - a c \kappa \left( {e  \over \rho c_{\rm v}} \right)^4,
\end{equation}
in the case where the energy in the radiation is much greater than that in the 
gas. In figure \ref{fig:ts51}, the solid line is this analytic solution, 
plotted both for the cases where $T_{\rm g}$ increases and decreases. The
crosses are the results of the SPH code using an implicit timestep that is 
set to  
the larger of $10^{-14}$ s or five percent of the time elapsed. The squares
are similar, but with a timestep being the greater of $10^{-11}$ s or
five percent of the time elapsed.
As can be seen, the match between the analytic solution and the solutions 
given by the SPH code is excellent.

\subsection{Propagating radiation in optically thin media}

\begin{figure}
\centerline{\psfig{figure=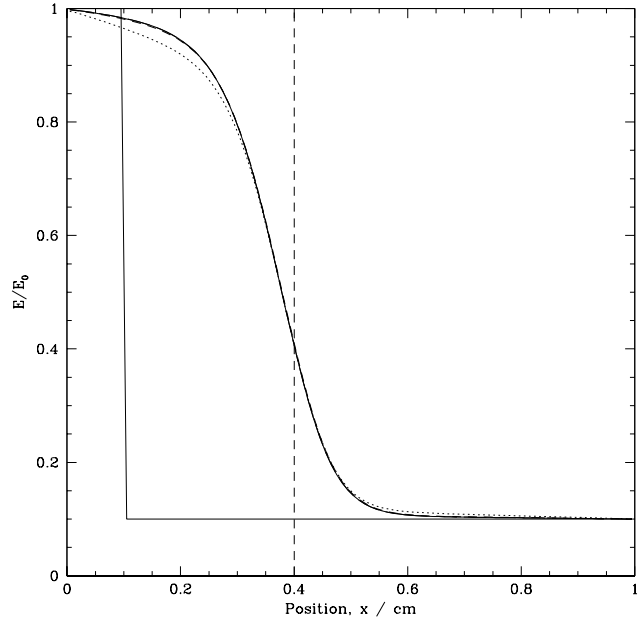,width=9.0truecm}}
\caption{\label{fig:ts55} 
The propagation of a radiation pulse across a uniform medium. The
time is $t=10^{-11}$~s. The vertical dashed line shows the expected position
of the pulse based on the speed of light. The results are essentially
independent of the size of the implicit timestep used.  Results are
given for implicit timesteps equal to (solid line),
ten times (dotted line), and one hundred times (dashed line) the explicit 
timestep.  The dot-dashed line gives the result using a single implicit 
step of $10^{-11}$~s.  The results for the explicit code are identical to that
of the implicit code using the explicit timestep, and hence are not plotted.
The initial conditions are also shown (solid line).}

\end{figure}

In the standard diffusion approximation, radiation can, in optically thin
regions, approach infinite velocity. This is unphysical, and the flux limiter
has been introduced to limit the diffusion of radiation to the speed of light. 
To examine how well our code limits the speed of the radiation, a one 
centimetre long one-dimensional
box is filled with 100 equally spaced SPH particles, with $E= 10^{-2}$ erg
cm$^{-3}$ ($\xi = 0.4$ erg g$^{-1}$), $\rho = 0.025$ g cm$^{-3}$ and
 $ \kappa=0.4$ cm$^2$ g$^{-1}$.  Initially, the radiation and gas are
in thermal equilibrium.

At the start of the simulation, the radiation energy density for the leftmost
ten particles was changed to $E=0.1$ erg cm$^{-3}$($\xi = 4$ erg g$^{-1}$) 
 and the resulting
radiation front was allowed to propagate across the region. 
The ghost particles were reflective except in specific radiation energy
$\xi$,  which was fixed equal to $\xi=4$ erg g$^{-1}$ at the left hand 
boundary and $\xi=0.4$ erg g$^{-1}$ at the right. The implicit code was used
with various timesteps ranging from an explicit timestep to a single
implicit step lasting $10^{-11}$~s. The results are shown in figure 
\ref{fig:ts55}. The result for the explicit code lies on
top of the result for the implicit code using an explicit timestep.

As can be seen, the radiation pulse propagates at the correct speed,
even using a single implicit timestep that is more than $10^4$ times 
the explicit timestep. The front is smoothed out
in a manner similar to the results of \citet{TS2001}; both methods are
quite diffusive in this situation.  We note \citet{TS2001} began with
an initial discontinuity in the radiation energy density of 21 orders
of magnitude.  We found that as the discontinuity increases in magnitude,
the radiation pulse propagates more slowly with SPH due smoothing of the 
gradient of $E$ and the consequent error in the value of the flux-limiter.

\subsection{Optically-thick (adiabatic) and optically-thin (isothermal) shocks}

A shock-tube test was set up to investigate the way the code simulated 
optically-thin and optically-thick regimes and the transition between them.
In the limit of high optical depth, the gas cannot cool because the 
radiation is trapped within the gas; thus the shock is adiabatic.  An 
optically-thin shock, on the other hand, is able to efficiently radiate 
away the thermal energy and thus behaves as an isothermal shock.  In these
shock tests, the gas and radiation are highly coupled and, thus, their 
temperatures are equal.

A domain $2 \times 10^{15}$ cm long extending from $x = -1 \times 10^{15}$ 
to $1 \times 10^{15} $ cm was set up, with an initial
density of $\rho = 10^{-10}$ g cm$^{-3}$, and the temperatures of the 
gas and radiation were initially $1500 $~K.  One
hundred particles were equally spaced in the domain, with those with negative
$x$ having a velocity equal to the adiabatic ($\gamma = 5/3$) sound speed
$v_0=c_{\rm s}= 3.2\times 10^5$ cm s$^{-1}$, and those with
positive $x$ travelling at the same speed in the opposite direction. 
The two flows impact at the origin, and a shock forms. 
Opacities of $\kappa = 40, 0.4,  4.0 \times 10^{-3} $ 
and $4.0 \times 10^{-5}$ cm$^2$ g$^{-1}$ were used to follow the transition
from adiabatic to isothermal behaviour.  Ghost particles were placed outside
the boundaries and maintain the initial quantities of their respective 
real particles. The boundaries moved inwards with the same velocities as the
initial velocities of the two streams.

\begin{figure}
\centerline{\psfig{figure=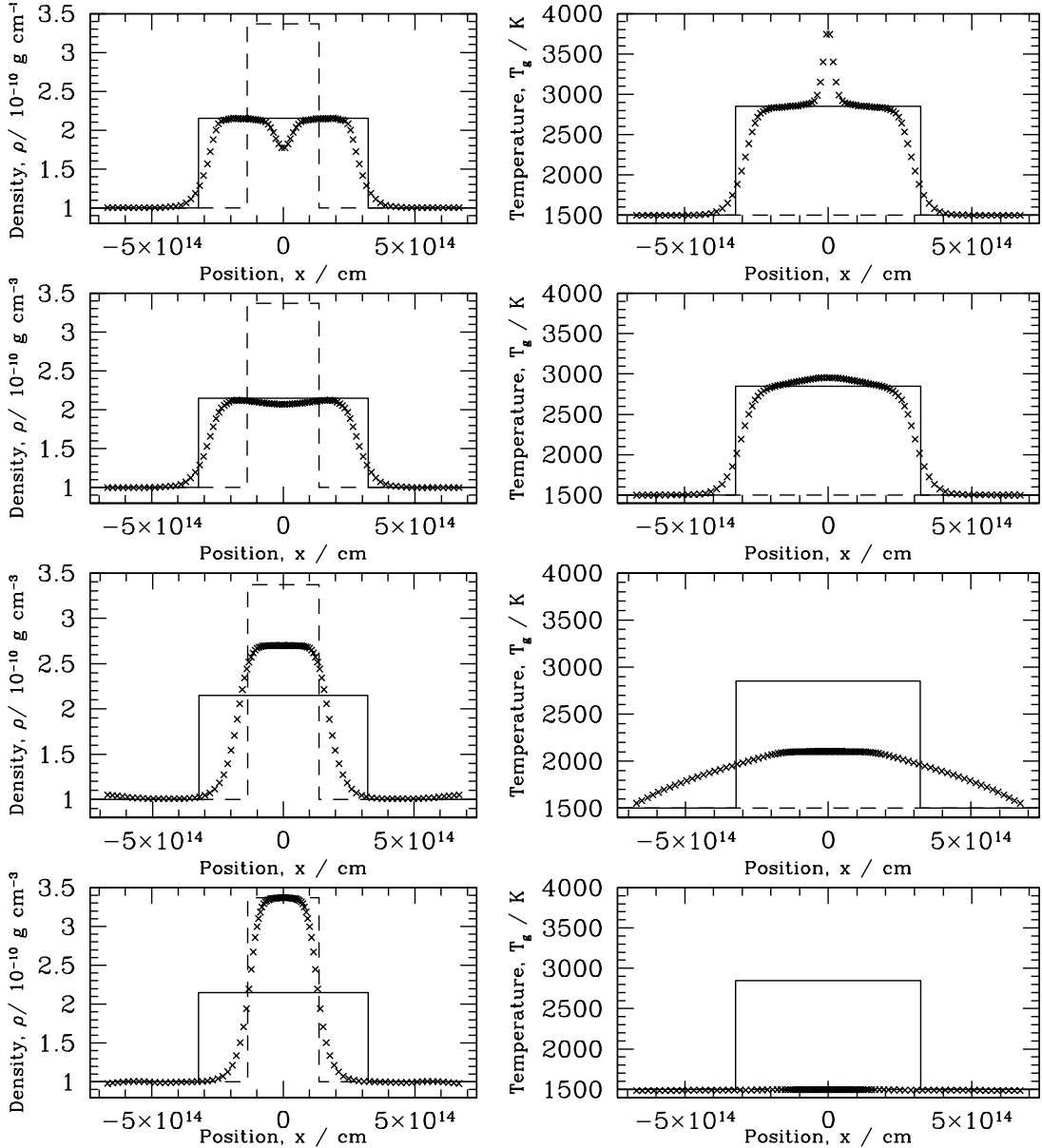,width=15.0truecm}}
\caption{\label{fig:isoadia} 
A set of shocks with differing opacity at time $t=1.0\times 10^9$~s. Density
is on the left, and gas temperature on the right. The crosses are the SPH
results; the solid line gives the analytic solution for an adiabatic shock, and
the dashed line for an isothermal shock. The opacities are (top) $\kappa = 
40$, (upper middle) $\kappa = 0.4 $, (lower middle)  $\kappa = 
4.0 \times 10^{-3}$ and (bottom) $\kappa = 4.0 \times 10^{-5}$ cm$^2$ g$^{-1}$.
As the opacity is decreased, the shocks transition from adiabatic to
isothermal behaviour.}
\end{figure}

The adiabatic and isothermal limits can be solved analytically 
(e.g. \citealp{Zeldo}).  The shock speed is given by
\begin{equation}
D = \frac{ ( \gamma_{\rm eff} - 3) + \sqrt{ (\gamma_{\rm eff} +1 )^2 v_0^2 + 16 \gamma_{\rm eff} }}{4},
\end{equation}
where $\gamma_{\rm eff}=5/3$ for the adiabatic case, and $\gamma_{\rm eff}=1$ 
for the isothermal case.  The ratio of the final to the initial density 
is given by 
\begin{equation}
\frac{\rho_1}{\rho_0} = 1 + {v_0 \over D},
\end{equation}
and, for the adiabatic shock, the ratio of the final to the initial temperature
is given by
\begin{equation}
\frac{T_1}{T_0} = \frac{\rho_1}{\rho_0} + \frac{v_0 D \rho_0}{p_0}.
\end{equation}
These analytic solutions are shown by the solid and dashed lines in figure
\ref{fig:isoadia}.  In the figure, the opacity decreases from top to bottom
showing the transition from optically-thick (adiabatic) to optically-thin 
(isothermal) behaviour.  The extremes are in good agreement with their
respective analytic adiabatic and isothermal solutions.  Note that the
spike in thermal energy near the origin and the corresponding reduction 
in density for the optically-thick case (due to `wall-heating') is softened
by the radiation transport that occurs in the intermediate opacity
calculation with $\kappa = 0.4$.

Where possible, comparison shows that the results from a fully explicit 
calculation are virtually identical to those obtained implicitly.  However,
as the opacity is increased, the implicit code offers an enormous 
improvement in speed because the explicit timestep is set by the interaction
term between the radiation and gas (equations \ref{tsint1} and \ref{tsint2})
which is inversely proportional to the opacity.  
With an opacity of $\kappa = 4.0\times 10^{-5}$,
the codes are similar in speed.  However, the implicit code is $\sim 100$
times faster than the explicit when $\kappa = 4.0\times 10^{-3}$, and 
$\sim 10^4$ times faster when $\kappa = 0.4$.  For $\kappa = 40$,
we could not run an explicit calculation, but the implicit code is likely
to be $\sim 10^6$ times faster.
%4e-6 changed to 4e-5
%4e3 changed to 4e1
% And times updated too

\subsection{Sub- and super-critical shocks}
\label{sec:radshocks}

\begin{figure}
\centerline{\psfig{figure=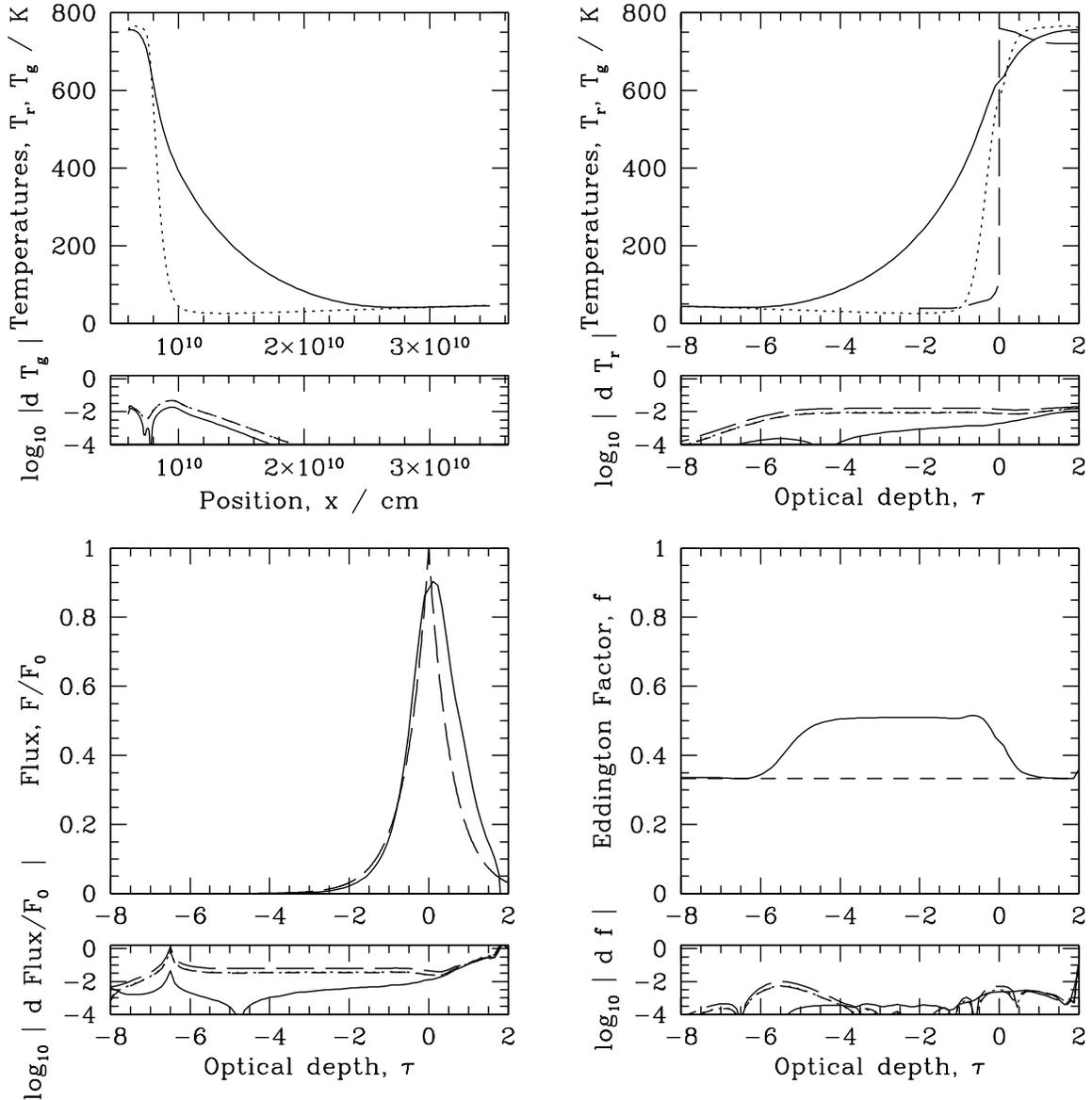,width=16.0truecm}}
\caption{\label{fig:subshock}
The sub-critical shock, with piston velocity $6 \times 10^5$ cm s$^{-1}$ and
100 particles. The 
large panels show the results using the explicit code. The top panels show 
radiation (solid line) and gas (dotted line) temperatures.  The bottom 
left panel shows the normalised flux and bottom right panel the Eddington 
factor. The long-dashed lines give the analytic solutions for the gas 
temperature and normalised flux.  An Eddington factor of 1/3 is also 
indicated for reference (short-dashed line, lower right panel).  
The subpanels plot the logarithm of the difference between the results 
using the explicit code and 
the implicit code (see the main text).  The implicit code was run with 
timesteps of 1 (solid line), 10 (short-dashed line), and 100 (dotted line)
times the explicit timestep and with a hydrodynamical timestep (long dashes).}
\end{figure}
 
\begin{figure}
\centerline{\psfig{figure=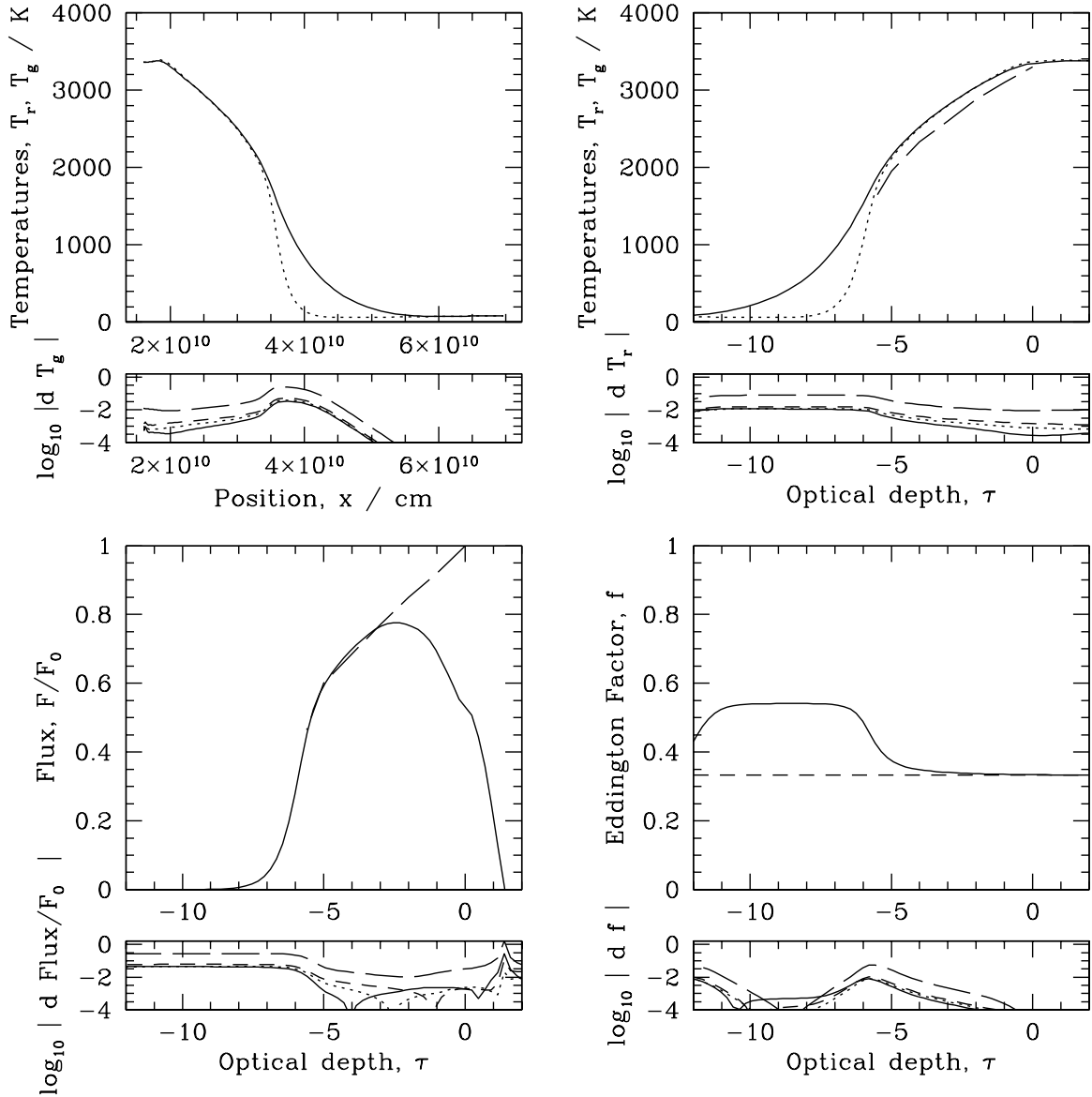,width=16.0truecm}}
\caption{\label{fig:supershock}
The super-critical shock, with piston velocity $1.6 \times 10^6$ cm s$^{-1}$ 
and
100 particles.  This shock is strong enough for radiation from the shock 
to preheat the gas upstream.  See figure \ref{fig:subshock} for more details.}
\end{figure}

\begin{figure}
\centerline{\psfig{figure=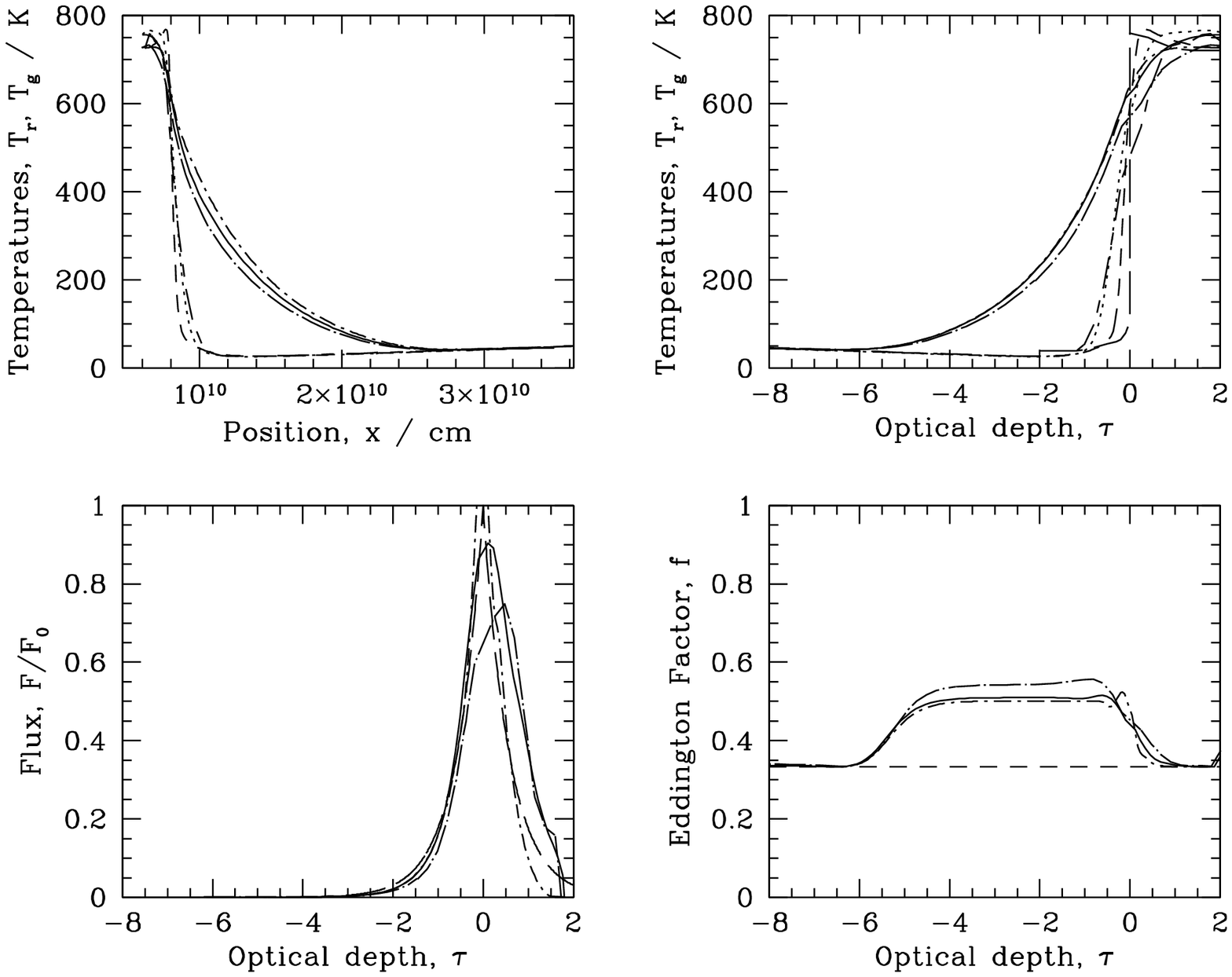,width=16.0truecm}}
\caption{\label{fig:subresol}
The sub-critical shock with differing resolutions. For 50 particles,
the dot-long-dashed lines give the radiation temperature, the flux, and the
Eddington factor while the short-dash-long-dashed lines give the gas 
temperature.  For 100 particles, the solid lines give the radiation 
temperature, the flux, and the Eddington factor while the dotted lines 
give the gas temperature.  For 250 particles, the dot-short-dashed 
lines give the radiation temperature, the flux, and the
Eddington factor while the short-dashed lines give the gas 
temperature.  The analytic solution is given by the long-dashed lines.}
\end{figure}
 
\begin{figure}
\centerline{\psfig{figure=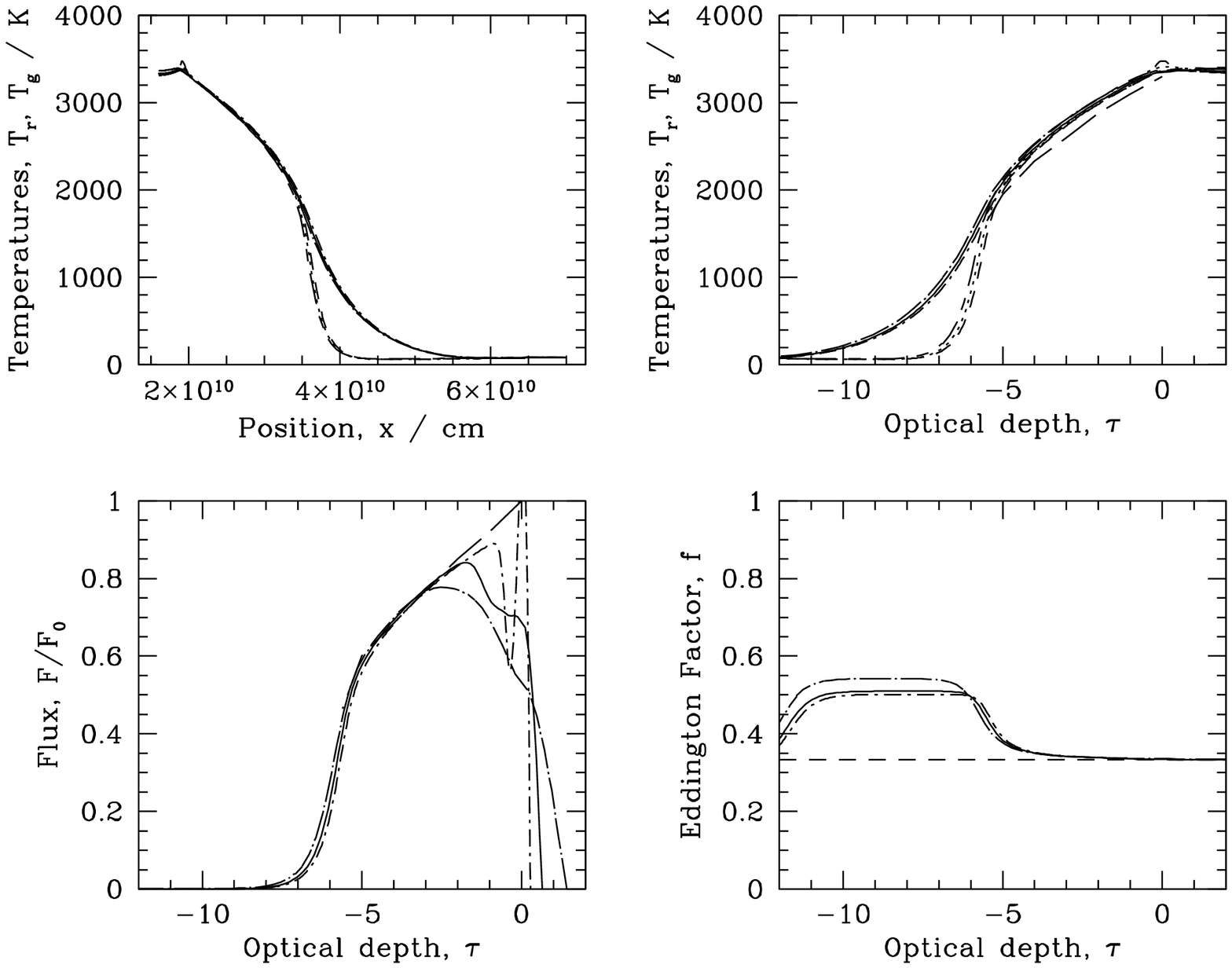,width=16.0truecm}}
\caption{\label{fig:superresol}
The super-critical shock with differing resolutions. For 100 particles,
the dot-long-dashed lines give the radiation temperature, the flux, and the
Eddington factor while the short-dash-long-dashed lines give the gas 
temperature.  For 200 particles, the solid lines give the radiation 
temperature, the flux, and the Eddington factor while the dotted lines 
give the gas temperature. For 500 particles, the dot-short-dashed 
lines give the radiation temperature, the flux, and the
Eddington factor while the short-dashed lines give the gas 
temperature. The analytic solution is given by the 
long-dashed lines.  }
\end{figure}
 
\begin{figure}
\centerline{\psfig{figure=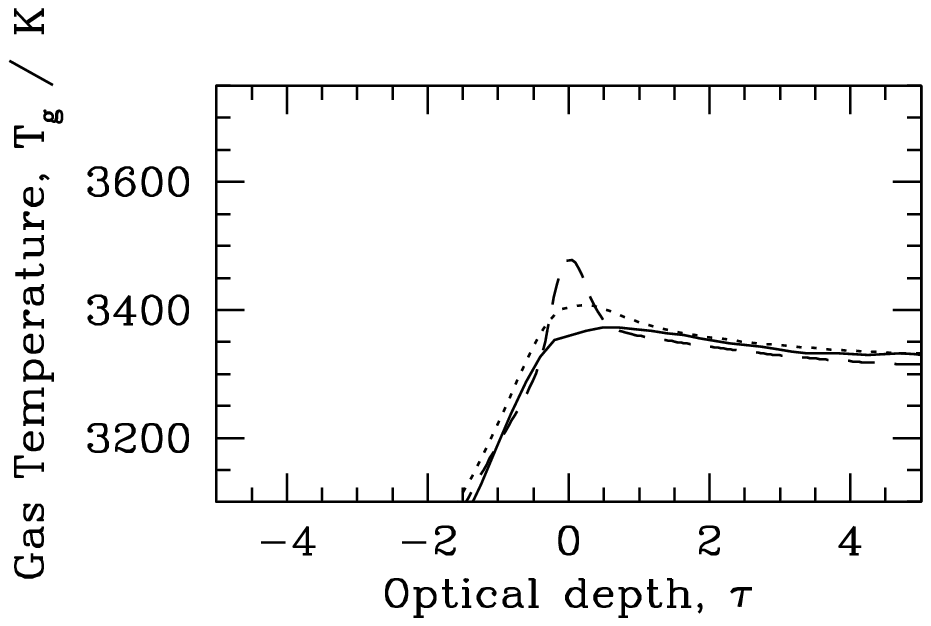,width=12.0truecm}}
\caption{\label{fig:spike} 
A zoom-in of the spike for the supercritical
shock. We plot the results with 100 (solid line), 200 (dotted
line) and 500 (dashed line) particles. As expected, the spike is more 
obvious with higher resolution. }
\end{figure}

A supercritical shock occurs when the photons generated by a shock have 
sufficient energy 
to preheat the material upstream. The characteristic temperature profile of
a supercritical shock is where the temperature on either side of the shock
is similar, rather than the downstream temperature being much higher than
the upstream, as occurs in a subcritical shock \citep[see][for more 
details]{Zeldo}. 

The initial conditions of this problem are those of \citet*{SGM1999} and
\citet{TS2001}. A gas with opacity 
$\kappa = 0.4 $ cm$^2$ g$^{-1}$, 
uniform density $\rho=7.78 \times 10^{-10}$ g cm$^{-3}$,  mean molecular
weight $\mu = 0.5$ and $\gamma = \frac{5}{3}$ is set up with $\xi$ and $u$
in equilibrium, with a temperature gradient of $T = 10 + \left( 75 x 
/ 7 \times 10^{10} \right)$ K. Initially the particles are
equally spaced between $x=0$ and $x= 7 \times 10^{10} $ cm for the
supercritical shock, and between $x=0$ and $x= 3.5 \times 10^{10} $ cm for
the subcritical shock. At time $t=0$ a
piston starts to move into the fluid from the left-hand boundary (simulated
by moving the location of the 
boundary). For the subcritical shock the piston velocity is
$v_{\rm p} = 6$ km s$^{-1}$, and for the supercritical shock $v_{\rm p}
 = 16$km s$^{-1}$, as
per \citet{SGM1999}. The ghost particles are reflective in the frame of
reference of the boundary. Artificial viscosity parameters 
$\alpha_{\rm v} = 2$ and $\beta_{\rm v} = 4$ were used to 
smooth out oscillations.

The results of calculations using 100 particles
are shown in figure \ref{fig:subshock} for the sub-critical shock
and in figure \ref{fig:supershock} for the super-critical shock. The 
top left panel for each shows the temperature of the radiation field (solid
line) and the gas (dotted line) against position, and the top right shows the
same quantities against optical depth $\tau$, with $\tau=0$ set at the
shock front (measured from the density distribution). The bottom left panel 
shows normalised flux, and the bottom right the value of the
Eddington factor.  The analytic solutions discussed by \citet{SGM1999} and
\citet{Zeldo} for the temperatures and fluxes of the shocks are
shown with long-dashed lines.
Figures \ref{fig:subshock} and \ref{fig:supershock}
are plotted using the explicit code.  In subpanels beneath the main panels,
we compare the results from the implicit code with the explicit results.
Calculations were performed using the implicit code with timesteps of 
1, 10, and 100 times the explicit timestep and with a timestep limited only
by the hydrodynamical timestep criteria (equations \ref{tshydro1} and 
\ref{tshydro2}).  In the subpanels, we plot the differences of the implicit
results with respect to the explicit results.  We divide the difference 
between the implicit and explicit values by the explicit value to obtain a
fractional error and take the logarithm of the absolute value of this 
fraction.  Thus, a difference of $-2$ on the subpanels corresponds to 
an error of 1 percent with respect to the explicit result. 

Figures \ref{fig:subresol}, \ref{fig:superresol} and \ref{fig:spike} 
show the effect on the
shocks of changing the resolutions.  For the sub-critical shock, the results
converge towards the analytic solutions for both the gas temperature and
the flux as the particle number is increased.    
Note that our highest resolution case has the same number of particles 
per unit length as \citet{TS2001} have grid cells.  Increasing the resolution
of the super-critical shock has less of an effect on the results, although
the flux in the vicinity of the shock improves, and the spike in gas
temperature at the location of the shock is better
resolved (c.f. figure \ref{fig:spike}).

Both the explicit and implicit codes model the sub-critical and super-critical
shocks well.  With similar resolution, our results are in good agreement with
those of \citet{TS2001}, although we note that \citet{TS2001} only used 
timesteps as large as ten radiation diffusion times.
However, we found that in these tests, our implicit code offers
little advantage over the explicit code.  With the super-critical shock
and an implicit timestep 10 times longer than the explicit timestep, the
speed is similar, but using a hydrodynamic timestep the code is roughly
10 times slower.  For the sub-critical shock, the situation is worse.
Running with an implicit timestep 100 times longer than the explicit 
timestep takes $\approx 60$ times longer and implicit calculation using 
a hydrodynamical timestep takes $\approx 400$ times longer than an 
explicit calculation.  The decrease in speed is due to the large number
of sweeps required by the implicit integration scheme in order to achieve
the required accuracy.  Increasing the tolerance above $10^{-3}$ results
in quicker calculations, roughly inversely proportional to the tolerance,
however the solution is less accurate and both the temperatures of the
gas and radiation are overestimated.

\subsection{Radiation-dominated shock}

In material of high optical depth the radiation generated in a shock cannot
diffuse away at a high rate, and so the radiation becomes confined
in a thin region adjacent to the shock.  \citet{TS2001} 
performed a calculation that tests whether the shock thickness is what one 
would expect in these
circumstances.  An extremely high Mach number shock (Mach number of 658) 
is set up, with the gas on the left having an initial density of 
$\rho = 0.01$ g cm$^{-3}$ (set by using different mass particles), 
opacity $\kappa=0.4$ cm$^2$ g$^{-1}$, temperature $T_{\rm r} = T_{\rm g} = 
10^4$ K, and speed $10^9$ cm s$^{-1}$. The gas on the right has density 
$\rho = 0.0685847$ g cm$^{-3}$, opacity $\kappa=0.4$ cm$^2$ g$^{-1}$, 
temperature $T_{\rm r} = T_{\rm g} = 4.239 \times 10^7$ K, and
speed $1.458 \times 10^8$ cm s$^{-1}$. The locations of the boundaries 
move with the same speed as their respective particles, and the properties 
of the ghost particles
outside these boundaries are fixed at their initial values.
We use the implicit code with a timestep 100 times larger than the explicit 
timestep.  1500 particles are equally spaced over a
domain extending from $x=-6 \times 10^5$ cm to $x=1.5 \times 10^5$ cm 
initially, with the discontinuity at $x = 0.5 \times 10^5$ cm.  
The location of the shock should be fixed in this frame, although 
individual particles flow through the shock.

After a period where a transient feature forms at the shock front 
and drifts downstream with the flow, a stable shock is established. Its 
thickness is expected to be roughly equal to the distance 
$l = { c \lambda / \kappa \rho u_1}$, 
where $u_1$ is the speed of material flowing into the shock front.

\begin{figure}
\centerline{\psfig{figure=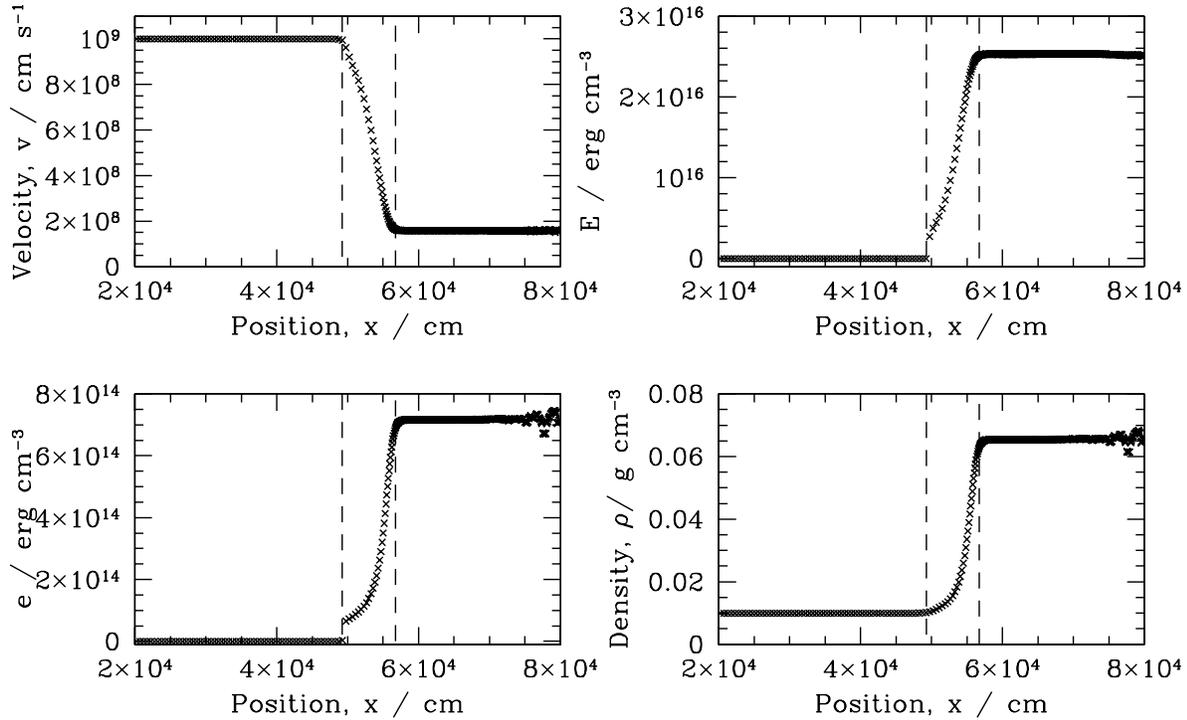,width=16.0truecm}}
\caption{\label{fig:radthick} The radiation-dominated shock at time 
$t= 5 \times 10^{-4}$~s.  We plot the velocity, radiation and gas energy
densities and gas density versus position.  The vertical dashed lines 
show the expected shock thickness.  Gas flows into the shock from the left.
The after effects of the transient that occurs at the start of the
calculation can be seen at the far right of the plots.    }
\end{figure}
 
Figure \ref{fig:radthick} shows the results at a
time $t= 5 \times 10^{-4}$~s, of the radiation energy density
$E$, gas energy density $e$, velocity $v$, and density $\rho$ versus
position.
The vertical dashed lines indicate the expected shock thickness and
the SPH results are in good agreement.   The after-effects of the transient
moving downstream can be seen on the right of the plots of density
and gas energy density. The explicit code takes roughly a factor of ten
times longer than the implicit code for this test.

\section{Conclusions}

In this paper, we have presented a method for including radiative transfer 
in the flux-limited diffusion approximation in a smoothed 
particle hydrodynamics (SPH) code.  The energy equations may be integrated
either explicitly or implicitly.  By integrating the energy equations
implicitly the short timesteps due to the rapid propagation of radiation 
can be avoided and timesteps limited only by hydrodynamical processes can
be used. In problems where the temperatures of the gas and radiation 
are equal, the implicit code loses little in accuracy but may be many
orders of magnitude faster than an explicit calculation.  However, 
in cases where the two temperatures are not equal the implicit scheme 
may actually be slower due to the large number of sweeps required to
obtain an accurate solution.  

Overall, the SPH code performs as accurately as the ZEUS code with
radiative transfer in the flux-limited diffusion approximation as
described by \citet{TS2001}.  Although the code and test calculations 
presented here are one-dimensional, it is straightforward to implement the
radiation hydrodynamical equations in a three-dimensional SPH code
due to the fact that the SPH method is structured around interactions
between pairs of particles.  
 In going from one to two or three dimensions,
the general method is the same. The main difference is that the radiation
pressure term in the evolution equation for the specific radiation energy
 involves a more complicated tensor equation.
Aside from this, the major 
limitation associated with increasing the number of dimensions is
that more particles are required to obtain the same resolution, thus more 
computational effort is required. Three-dimensional calculations will
be presented in a subsequent paper. We plan to
use this method to investigate the effects of radiative transfer on 
star formation. Other applications include modelling phenomena such as
accretion discs and radiation-dominated flows.

\section*{Acknowledgments}

The authors are grateful to Neal Turner and Jim Stone for discussions 
regarding test calculations for radiation hydrodynamics and
critical readings of the manuscript, and Charles D. H. 
Williams for discussions on implicit integration techniques.  SCW acknowledges 
support from a PPARC postgraduate studentship.

\appendix
\section{The analytic solution of a quartic equation}
\label{sec:quartsolve}

To solve the quartic equation $x^4 + a_3 x^3 + a_2 x^2 +a_1 x + a_0 =0$, it is
 necessary to first find the real root $y_1$ of the resolvent cubic
\begin{equation}
y^3 - a_2 y^2 + \left( a_1 a_3 - 4 a_0 \right) y + \left( 4 a_2 a_0 - a_1^2 - a_3^2 a_0 \right) = 0~,
\end{equation}
either analytically or numerically. The algorithm for a real root (of which
there must be at least one) can be obtained from a mathematical manipulation
package and output as computer code to be copied into a program. 
The four roots of the quartic are then given by the four solutions of

\begin{equation}
z^2 + \left[ {a_3 \over 2} \mp \left( {a_3^2 \over 4} + y_1 - a_2 \right)^{1 
\over 2} \right] z + \left[ { y_1 \over 2 } \mp \left({ y_1 \over 2} \right)^2
- a_0 \right]^{1 \over 2} = 0~,
\end{equation}
\citep{AbramS}. If some of the coefficients are large and of roughly 
equal size it may be necessary to Taylor expand the relevant portion of the
equation to maintain accuracy.

\bibliographystyle{mn2e}
\bibliography{paper}

\begin{thebibliography}{}

\bibitem[\protect\citeauthoryear{{Abramowitz} \& {Stegun}}{{Abramowitz} \&
  {Stegun}}{1972}]{AbramS}
{Abramowitz} M.,  {Stegun} I.~A.,  1972, Handbook of Mathematical Functions
  with Formulas, Graphs, and Mathematical Tables, 10th printing.
Dover, New York

\bibitem[\protect\citeauthoryear{{{Balsara}, D.}}{{{Balsara},
  D.}}{1989}]{Balsara}
{{Balsara}, D.} 1989, Doctoral ~thesis, University of Illinois

\bibitem[\protect\citeauthoryear{{Benz}, {Cameron}, {Press} \& {Bowers}}{{Benz}
  et~al.}{1990}]{BCPB1990}
{Benz} W.,  {Cameron} A.~G.~W.,  {Press} W.~H.,    {Bowers} R.~L.,  1990, ApJ,
  348, 647

\bibitem[\protect\citeauthoryear{{Black} \& {Bodenheimer}}{{Black} \&
  {Bodenheimer}}{1975}]{BB1975}
{Black} D.~C.,  {Bodenheimer} P.,  1975, Ap. J, 199, 619

\bibitem[\protect\citeauthoryear{{Brookshaw}}{{Brookshaw}}{1985}]{B1985}
{Brookshaw} L.,  1985, Proceedings of the Astronomical Society of Australia, 6,
  207

\bibitem[\protect\citeauthoryear{{Brookshaw}}{{Brookshaw}}{1986}]{B1986}
{Brookshaw} L.,  1986, Proceedings of the Astronomical Society of Australia, 6,
  461

\bibitem[\protect\citeauthoryear{{Cha} \& {Whitworth}}{{Cha} \&
  {Whitworth}}{2003}]{CW2003}
{Cha} S.-H.,  {Whitworth} A.~P.,  2003, MNRAS, 340, 73

\bibitem[\protect\citeauthoryear{{{Cha}, S.-~H.}}{{{Cha}, S.-~H.}}{2002}]{Cha}
{{Cha}, S.-~H.} 2002, Doctoral ~thesis, University of Cardiff

\bibitem[\protect\citeauthoryear{{Cleary} \& {Monaghan}}{{Cleary} \&
  {Monaghan}}{1999}]{CM1999}
{Cleary} P.~W.,  {Monaghan} J.~J.,  1999, J. Comp. Phys., 148, 227

\bibitem[\protect\citeauthoryear{{Dahlquist}}{{Dahlquist}}{1963}]{D1963}
{Dahlquist} G.,  1963, Nordisk Tidskrift for Informations Behandling, 3, 27

\bibitem[\protect\citeauthoryear{{Evrard}}{{Evrard}}{1988}]{E1988}
{Evrard} A.~E.,  1988, MNRAS, 235, 911

\bibitem[\protect\citeauthoryear{{Flebbe}, {Muenzel}, {Herold}, {Riffert} \&
  {Ruder}}{{Flebbe} et~al.}{1994}]{FMHRR1994}
{Flebbe} O.,  {Muenzel} S.,  {Herold} H.,  {Riffert} H.,    {Ruder} H.,  1994,
  ApJ, 431, 754

\bibitem[\protect\citeauthoryear{{Gingold} \& {Monaghan}}{{Gingold} \&
  {Monaghan}}{1977}]{GM1977}
{Gingold} R.~A.,  {Monaghan} J.~J.,  1977, MNRAS, 181, 375

\bibitem[\protect\citeauthoryear{{Gingold} \& {Monaghan}}{{Gingold} \&
  {Monaghan}}{1983}]{GM1983}
{Gingold} R.~A.,  {Monaghan} J.~J.,  1983, MNRAS, 204, 715

\bibitem[\protect\citeauthoryear{{Hernquist} \& {Katz}}{{Hernquist} \&
  {Katz}}{1989}]{HK1989}
{Hernquist} L.,  {Katz} N.,  1989, Ap. J. Suppl., 70, 419

\bibitem[\protect\citeauthoryear{{Inutsuka}}{{Inutsuka}}{1994}]{I1994}
{Inutsuka} S.,  1994, Memorie della Societa Astronomica Italiana, 65, 1027

\bibitem[\protect\citeauthoryear{{Inutsuka} \& {Imaeda}}{{Inutsuka} \&
  {Imaeda}}{2001}]{II2001}
{Inutsuka} S.-I.,  {Imaeda} Y.,  2001, Comp. Fluid. Dynam. J., 9, 316

\bibitem[\protect\citeauthoryear{{{Inutsuka}, S.-~I.}}{{{Inutsuka},
  S.-~I.}}{1998}]{I1999}
{{Inutsuka}, S.-~I.} 1998, {in Shoken M., Miyama Kohji Tomisaka T.H.., eds,
  Astrophysics and Space Science Library, Vol.240, Numerical Astrophysics,
  Proc. Int. Conf. on Numerical Astrophysics}.
Kluwer, Dordrecht, pp 367--374

\bibitem[\protect\citeauthoryear{{Jubelgas}, {Springel} \& {Dolag}}{{Jubelgas}
  et~al.}{2004}]{JSD2004}
{Jubelgas} M.,  {Springel} V.,    {Dolag} K.,  2004, astro-ph/0401456

\bibitem[\protect\citeauthoryear{{Levermore} \& {Pomraning}}{{Levermore} \&
  {Pomraning}}{1981}]{LP1981}
{Levermore} C.~D.,  {Pomraning} G.~C.,  1981, ApJ, 248, 321

\bibitem[\protect\citeauthoryear{{Lucy}}{{Lucy}}{1977}]{L1977}
{Lucy} L.~B.,  1977, AJ, 82, 1013

\bibitem[\protect\citeauthoryear{{Mihalas} \& {Mihalas}}{{Mihalas} \&
  {Mihalas}}{1984}]{MM1984}
{Mihalas} D.,  {Mihalas} B.~W.,  1984, {Foundations of Radiation
  Hydrodynamics}.
Oxford University Press

\bibitem[\protect\citeauthoryear{{Monaghan}}{{Monaghan}}{1992}]{M1992}
{Monaghan} J.~J.,  1992, Ann. Rev. Astron. Astrophys., 30, 543

\bibitem[\protect\citeauthoryear{{Monaghan}}{{Monaghan}}{1997}]{M1997}
{Monaghan} J.~J.,  1997, J. Comp. Phys., 138, 801

\bibitem[\protect\citeauthoryear{{Monaghan}}{{Monaghan}}{2002}]{M2002}
{Monaghan} J.~J.,  2002, MNRAS, 335, 843

\bibitem[\protect\citeauthoryear{{Morris} \& {Monaghan}}{{Morris} \&
  {Monaghan}}{1997}]{MM1997}
{Morris} J.~P.,  {Monaghan} J.~J.,  1997, J. Comp. Phys., 136, 41

\bibitem[\protect\citeauthoryear{{Oxley} \& {Woolfson}}{{Oxley} \&
  {Woolfson}}{2003}]{OW2003}
{Oxley} S.,  {Woolfson} M.~M.,  2003, MNRAS, 343, 900

\bibitem[\protect\citeauthoryear{{Price} \& {Monaghan}}{{Price} \&
  {Monaghan}}{2004a}]{PM2004a}
{Price} D.~J.,  {Monaghan} J.~J.,  2004a, MNRAS, 348, 123

\bibitem[\protect\citeauthoryear{{Price} \& {Monaghan}}{{Price} \&
  {Monaghan}}{2004b}]{PM2004b}
{Price} D.~J.,  {Monaghan} J.~J.,  2004b, MNRAS, 348, 139

\bibitem[\protect\citeauthoryear{{Sincell}, {Gehmeyr} \& {Mihalas}}{{Sincell}
  et~al.}{1999}]{SGM1999}
{Sincell} M.~W.,  {Gehmeyr} M.,    {Mihalas} D.,  1999, Shock Waves, 9, 391

\bibitem[\protect\citeauthoryear{{Springel} \& {Hernquist}}{{Springel} \&
  {Hernquist}}{2002}]{SH2002}
{Springel} V.,  {Hernquist} L.,  2002, MNRAS, 333, 649

\bibitem[\protect\citeauthoryear{{Turner} \& {Stone}}{{Turner} \&
  {Stone}}{2001}]{TS2001}
{Turner} N.~J.,  {Stone} J.~M.,  2001, ApJS, 135, 95

\bibitem[\protect\citeauthoryear{{Watkins}, {Bhattal}, {Francis}, {Turner} \&
  {Whitworth}}{{Watkins} et~al.}{1996}]{WBFTW1996}
{Watkins} S.~J.,  {Bhattal} A.~S.,  {Francis} N.,  {Turner} J.~A.,
  {Whitworth} A.~P.,  1996, A\&AS, 119, 177

\bibitem[\protect\citeauthoryear{{Wood}}{{Wood}}{1981}]{W1981}
{Wood} D.,  1981, MNRAS, 194, 201

\bibitem[\protect\citeauthoryear{{Zel'dovich} \& {Raizer}}{{Zel'dovich} \&
  {Raizer}}{2002}]{Zeldo}
{Zel'dovich} Y.~B.,  {Raizer} Y.~P.,  2002, Physics of Shock Waves and
  High-Temperature Hydrodynamic Phenomena, edited Hayes and Probstein.
Dover, New York

\end{thebibliography}

\end{document}